\documentclass[a4paper,11pt]{article}
\pdfoutput=1
\usepackage{jcappub}
\usepackage{mathtools}
\graphicspath{{plots/}}

\newcommand{\gs}{g_\text{eff}}
\newcommand{\gss}{h_\text{eff}}
\newcommand{\Trh}{T_\text{rh}}

\newcommand{\Tmax}{T_\text{max}}
\newcommand{\Tfo}{T_\text{fo}}
\newcommand{\sv}{\langle\sigma v\rangle}
\newcommand{\rR}{\rho_R}
\newcommand{\rp}{\rho_\phi}

\title{WIMPs as FIMPs: the Inert Doublet Model and low-temperature reheating}
\author[a]{Geneviève Bélanger,}
\author[b]{Nicolás Bernal,}
\author[c]{Andreas Goudelis,}
\author[d]{\\Alexander Pukhov,}
\author[c]{Thomas Reggio}

\affiliation[a]{LAPTh, CNRS, Université Savoie Mont-Blanc\\
9 Chemin de Bellevue, 74940 Annecy, France}
\affiliation[b]{New York University Abu Dhabi\\
PO Box 129188, Saadiyat Island, Abu Dhabi, United Arab Emirates}
\affiliation[c]{Laboratoire de Physique de Clermont Auvergne (UMR 6533), CNRS/IN2P3\\
Univ. Clermont, Auvergne, 4 Av. Blaise Pascal, F-63178 Aubière Cedex, France}
\affiliation[d]{Skobeltsyn Institute of Nuclear Physics, Moscow State University\\
Moscow 119992, Russia}

\emailAdd{belanger@lapth.cnrs.fr}
\emailAdd{nicolas.bernal@nyu.edu}
\emailAdd{andreas.goudelis@clermont.in2p3.fr}
\emailAdd{alexander.pukhov@gmail.com}
\emailAdd{thomas.reggio@clermont.in2p3.fr}

\abstract{Weakly Interacting Massive Particles (WIMPs) are conventionally assumed to acquire their dark matter relic abundance through thermal freeze-out. However, this association relies on assumptions concerning the thermal history of the early Universe. We revisit the Inert Doublet Model (IDM), a representative WIMP framework in which dark matter interacts with the Standard Model through electroweak gauge interactions, even when its Higgs-portal coupling vanishes. We demonstrate that sufficiently low reheating temperatures can prevent chemical equilibrium, allowing a genuine WIMP to acquire its relic abundance through freeze-in rather than freeze-out. Considering both instantaneous and finite-duration reheating, we identify different production histories, including freeze-out during radiation domination, freeze-out during reheating, and freeze-in with potentially significant annihilation backreactions. These scenarios open regions of the IDM parameter space that would otherwise be excluded by dark matter overabundance. Our results demonstrate that WIMPs can behave as FIMPs, depending on the thermal history of the Universe.}
\begin{document}
\begin{flushright}
\end{flushright}
\maketitle

%%%%%%%%%%%%%%%%%%%%%%%%%%%%%%%%%%%%%%%%%%%%%%%%%%%%%%%%%%
\section{Introduction}
%%%%%%%%%%%%%%%%%%%%%%%%%%%%%%%%%%%%%%%%%%%%%%%%%%%%%%%%%%
The existence of dark matter (DM) constitutes one of the most compelling indications of physics beyond the Standard Model (SM) of particle physics. Despite the wealth of gravitational evidence supporting its existence, the microscopic nature of DM and the mechanism responsible for generating its cosmic abundance remain unknown. Over the past decades, numerous DM candidates have been proposed, typically associated with specific production mechanisms in the early Universe~\cite{Planck:2018vyg, Cirelli:2024ssz}. Among the most extensively studied possibilities are Weakly Interacting Massive Particles (WIMPs) and Feebly Interacting Massive Particles (FIMPs). In the conventional picture, WIMPs are assumed to reach chemical equilibrium with the SM thermal bath before their abundance is set by thermal freeze-out~\cite{Arcadi:2017kky, Roszkowski:2017nbc, Arcadi:2024ukq}. FIMPs, by contrast, never reach chemical equilibrium, and their relic abundance is gradually generated through freeze-in~\cite{McDonald:2001vt, Choi:2005vq, Kusenko:2006rh, Petraki:2007gq, Hall:2009bx, Elahi:2014fsa, Bernal:2017kxu}. Therefore, these two production mechanisms are commonly associated with relatively strong and feeble interactions with the SM, respectively.

As natural as this association may appear, it relies on assumptions concerning the thermal history of the early Universe. In particular, conventional WIMP relic-density calculations typically assume that the SM plasma reached temperatures sufficiently high for DM to attain chemical equilibrium, an assumption often referred to as the effectively infinite reheating temperature approximation. Moreover, the onset of radiation domination is frequently described within the instantaneous-reheating approximation, in which the energy density of the field responsible for reheating is converted into SM radiation over a negligibly short interval. Although these assumptions provide a simple framework for computing the DM abundance, the thermal history of the Universe before Big Bang Nucleosynthesis remains largely unconstrained~\cite{Allahverdi:2020bys, Batell:2024dsi}. In particular, the temperature at the onset of radiation domination need not have been much higher than the characteristic mass scales of DM production.

Relaxing these assumptions can substantially modify the connection between the microscopic properties of DM and its cosmological production mechanism. In particular, if the SM plasma never reaches sufficiently high temperatures, the thermal production of massive particles can become strongly Boltzmann suppressed, preventing chemical equilibrium from being established even when their interactions with the SM are relatively strong. Consequently, DM candidates conventionally associated with thermal freeze-out can instead acquire their relic abundance through freeze-in. In recent years, several studies have demonstrated that freeze-in can occur for interaction strengths substantially larger than those usually associated with FIMPs~\cite{Belanger:2018sti, Bernal:2022wck, Bhattiprolu:2022sdd, Cosme:2023xpa, Silva-Malpartida:2023yks, Cosme:2024ndc, Koivunen:2024vhr, Arcadi:2024wwg, Boddy:2024vgt, Silva-Malpartida:2024emu, Arcadi:2024obp, Lebedev:2024mbj, Bernal:2024ndy, Lee:2024wes, Belanger:2024yoj, Khan:2025keb, Bernal:2025osg, Bernal:2025szh, Bernal:2025qkj, Arias:2025tvd, Roy:2025moo, Bernal:2025fcl, Khan:2025kuh, Bernal:2026clv, Feiteira:2026qme, Belanger:2026ctm, Cerdeno:2026lgw, Bernal:2026vbg, C:2026bqd, Ghosh:2026mda, Bertou:2026osq, Goudelis:2026lyy, Gross:2026bhe, Bernal:2026mxg, Choudhury:2026aao}. Most investigations of this possibility have focused on DM candidates that are singlets under the SM gauge group, whose interactions with the visible sector are controlled by adjustable portal couplings. Its implications for conventional WIMP models, in which electroweak interactions are dictated by gauge invariance, have received comparatively less attention~\cite{Bernal:2026clv, Bernal:2026vbg}.

A particularly interesting framework in which to investigate this possibility is the Inert Doublet Model (IDM)~\cite{Deshpande:1977rw, Barbieri:2006dq, LopezHonorez:2006gr, LopezHonorez:2010tb, Goudelis:2013uca}, one of the simplest and most extensively studied extensions of the SM accommodating a viable DM candidate. The model introduces an additional scalar doublet, odd under an exact discrete $\mathbb{Z}_2$ symmetry, whose lightest neutral component is stable. Being part of an $SU(2)_L$ doublet, the DM candidate necessarily interacts with the SM through electroweak gauge interactions. Importantly, these interactions remain present even if
the model's parameters are chosen such that the tree-level Higgs-portal
coupling of the DM candidate vanishes. Moreover, its mass naturally lies at the electroweak scale or above in the parameter space of interest. The IDM DM candidate therefore possesses the defining characteristics of a WIMP in the particle-physics sense: a massive particle interacting with the SM through interactions of electroweak strength.

Under standard cosmological assumptions, these properties naturally lead to chemical equilibrium with the SM plasma and subsequent thermal freeze-out. However, as we demonstrate in this work, this connection is not inevitable. If the temperature of the SM plasma is sufficiently low, thermal production can be strongly suppressed, preventing chemical equilibrium despite the presence of electroweak gauge interactions. The very same particle can then acquire its relic abundance through freeze-in rather than freeze-out, without any modification of its microscopic interactions. In other words, a particle that is a WIMP from the perspective of its mass and couplings can behave as a FIMP from the perspective of its cosmological production mechanism. This is the central idea underlying the title of our work: WIMPs can effectively behave as FIMPs.

In this work, we systematically revisit the IDM relic-density predictions in scenarios featuring low-temperature reheating. We first consider the instantaneous-reheating approximation, in which radiation-dominated evolution begins at a temperature $\Trh$ that is not necessarily higher than the DM mass. We show that as $\Trh$ decreases, the conventional freeze-out solution can continuously evolve into a nonequilibrium production regime, potentially involving significant annihilation backreactions. This opens regions of the IDM parameter space that would otherwise be excluded by DM overabundance. We also investigate the resulting restrictions on the DM mass and the dependence of the viable parameter space on the inert-scalar mass splittings and Higgs-portal coupling.

We then extend our analysis beyond the instantaneous-reheating approximation, considering a finite period during which the field dominating the energy density of the Universe gradually decays into SM radiation. In this case, the plasma can reach a maximum temperature $\Tmax$ exceeding the temperature $\Trh$ associated with the transition to radiation domination. Dark matter production can therefore occur during reheating itself, in a cosmological background characterized by a modified expansion rate and ongoing entropy injection. We show that, depending on the hierarchy between the particle masses and the relevant cosmological temperatures, the observed abundance can arise through several distinct production histories, ranging from conventional freeze-out during radiation domination to freeze-out during reheating and freeze-in. In particular, nonequilibrium production can take place during the reheating epoch, with annihilation backreactions potentially affecting the final abundance.

The remainder of this paper is organized as follows. In Section~\ref{sec:cosmo}, we review the dynamics of reheating and the Boltzmann formalism governing DM production. In Section~\ref{sec:idm}, we introduce the IDM, summarize the relevant theoretical and experimental constraints, and revisit its relic-density predictions under standard cosmological assumptions. In Section~\ref{sec:results}, we investigate the impact of low-temperature reheating, first within the instantaneous-reheating approximation and subsequently allowing for a finite reheating period. Finally, in Section~\ref{sec:conclu}, we summarize our main findings and conclude.

%%%%%%%%%%%%%%%%%%%%%%%%%%%%%%%%%%%%%%%%%%%%%%%%%%%%%%%%%%
\section{Dark matter production during reheating} \label{sec:cosmo}
%%%%%%%%%%%%%%%%%%%%%%%%%%%%%%%%%%%%%%%%%%%%%%%%%%%%%%%%%%
As already mentioned in the Introduction, the standard picture of DM production in the early Universe rests on two assumptions: that the SM plasma reached a temperature higher than the mass scales at which DM is effectively produced (the ``effectively infinite'' reheating temperature assumption) and that this reheating temperature was also the maximal temperature ever attained by the plasma (the instantaneous reheating approximation). In this Section we discuss how these assumptions can be relaxed and introduce some formalism which will be employed in the following.

%%%%%%%%%%%%%%%%%%%%%%%%%%%%%%%%%%%%%%%%%%%%%%%%%%%%%%%%%%
\subsection{Reheating dynamics} \label{sec:reheatingdynamics}
%%%%%%%%%%%%%%%%%%%%%%%%%%%%%%%%%%%%%%%%%%%%%%%%%%%%%%%%%%
After the end of cosmic inflation, we consider an early epoch of inflaton domination, during which the energy density of the Universe is dominated by an inflaton condensate $\phi$ that oscillates in a quadratic potential while decaying into SM radiation with constant decay width $\Gamma_\phi$: the cosmic reheating era. The evolution of the inflaton and radiation energy densities, $\rp$ and $\rR$, is governed by the continuity equations and the first Friedmann equation
\begin{align}
    \dot{\rp} + 3\, H\, \rp &= -\Gamma_\phi\, \rp\,,  \label{eq:background1}\\
    \dot{\rR} + 4\, H\, \rR &= +\Gamma_\phi\, \rp\,,  \label{eq:background2}\\
    H^2 &= \frac{\rp + \rR}{3\, M_P^2}\,, \label{eq:background3}
\end{align}
where $H$ is the Hubble expansion rate and $M_P \simeq 2.4 \times 10^{18}$~GeV is the reduced Planck mass, and we have assumed that the decay products of the inflaton thermalize efficiently. As mentioned previously, in minimal scenarios $\phi$ plays the role of the inflaton. However, in general, $\phi$ can also be a generic massive field (\textit{i.e.} a modulus field) that dominates the total energy budget of the Universe while decaying into SM states. Equations~\eqref{eq:background1} and~\eqref{eq:background2} are solved from an initial scale factor $a = a_I$ with the initial conditions $\rR(a_I) \simeq 0$ and $\rp(a_I) = 3\, M_P^2\, H_I^2$, where $H_I \equiv H(a_I)$ corresponds to the initial Hubble scale.

At early times, as long as $\Gamma_\phi \ll H(a)$, the inflaton decays can be neglected, and therefore its energy density scales as nonrelativistic matter: $\rp(a) \simeq \rp(a_I)\, (a_I/a)^3$. However, inefficient inflaton decays source the SM radiation, giving
\begin{equation}
    \rR(a) \simeq \frac65\, M_P^2\, H_I\, \Gamma_\phi \left(\frac{a_I}{a}\right)^{3/2} \left[1 - \left(\frac{a_I}{a}\right)^{5/2}\right],
\end{equation}
which is valid as long as $\rp \gg \rR$. The SM energy density can be expressed as a function of the SM temperature $T$ as $\rR(T) = \frac{\pi^2}{30}\, \gs(T)\, T^4$, where $\gs(T)$ corresponds to the number of relativistic degrees of freedom contributing to $\rR$. After reaching its maximum value, and as long as the inflaton dominates the energy density, the temperature of the SM plasma approximately scales as
\begin{equation}
    T(a) \propto a^{-3/8}\,,
\end{equation}
up to variations of $\gs(T)$, with a scaling characteristic of an early-matter domination. We note that here we are assuming an instantaneous thermalization of the SM sector. In practice, the temperature first increases from $T(a_I) \simeq 0$ to a maximum value $\Tmax$ at $a/a_I = (8/3)^{2/5} \simeq 1.5$ and then drops monotonically~\cite{Giudice:2000ex}. Maximizing $T(a)$ analytically yields
\begin{equation}
    \Tmax^4 \simeq \frac{60}{\pi^2\, \gs(\Tmax)} \left(\frac38\right)^{8/5} M_P^2\, \Gamma_\phi\, H_I\,.
\end{equation}
This expression has been derived assuming $\Gamma_\phi \ll H_I$, such that inflaton depletion and the contribution of radiation to the Hubble expansion rate can be neglected until the maximum temperature is reached. For more general values of $\Gamma_\phi / H_I$, the maximum temperature is obtained by numerically solving Eqs.~\eqref{eq:background1} to~\eqref{eq:background3}.

Once $\Gamma_\phi \simeq H(\Trh)$, the inflaton decays efficiently, at a reheating temperature $\Trh$ corresponding to
\begin{equation}
    \Trh^2 \simeq \frac{3}{\pi}\, \sqrt{\frac{10}{\gs(\Trh)}}\, M_P\, \Gamma_\phi\,.
\end{equation}
In general, $\Trh$ is a proxy of the temperature at which the inflaton and radiation energy densities become equal. In this vein, in everything that follows, $\Trh$ will be determined by a numerical solution of the background equations~\eqref{eq:background1}-\eqref{eq:background3} and defined through $\rho_{\rm infl} = \rho_{\rm }$. After reheating is completed, the Universe enters the conventional radiation-dominated era, during which the comoving SM entropy is conserved and $T(a) \propto a^{-1}$, up to variations of the relativistic number of degrees of freedom $\gss(T)$ contributing to the SM entropy.

Together, there are two relevant temperature scales characterizing the cosmological evolution: the maximal temperature $\Tmax$ attained by the plasma and the reheating temperature $\Trh$. In the instantaneous reheating approximation, it is assumed that the inflaton energy density is converted into SM radiation instantaneously, so that the radiation-dominated evolution starts at $T = \Tmax = \Trh$. For the upcoming analysis, we numerically solve the system of Boltzmann Eqs.~\eqref{eq:background1} to~\eqref{eq:background3}.

%%%%%%%%%%%%%%%%%%%%%%%%%%%%%%%%%%%%%%%%%%%%%%%%%%%%%%%%%%
\subsection{Freeze-out versus freeze-in} \label{sec:boltzmann}
%%%%%%%%%%%%%%%%%%%%%%%%%%%%%%%%%%%%%%%%%%%%%%%%%%%%%%%%%%
In the case where DM is produced from the SM thermal bath through 2-to-2 reactions, the evolution of the DM number density $n$ is given by
\begin{equation} \label{eq:BE0}
    \dot{n} + 3\, H\, n = - \sv \left(n^2 - n_{\rm eq}^2\right),
\end{equation}
where $\sv$ corresponds to the thermally-averaged DM annihilation cross section, and $n_{\rm eq}(T)$ to its number density at equilibrium. In particular, we assume that there is no DM production from the direct decay of the inflaton. Equation~\eqref{eq:BE0} describes the evolution of a single DM species undergoing annihilations into SM particles. More generally, when several dark-sector particles participate in the production and depletion processes, their abundances are governed by a system of coupled Boltzmann equations, including annihilations, coannihilations, conversions, and decays. This is particularly relevant for models involving multiple states with nearly degenerate masses. In our numerical analysis, we account for all these processes by evolving the relevant out-of-equilibrium species.

When the comoving SM entropy is conserved, as in the conventional radiation-dominated era after reheating, it is convenient to rewrite Eq.~\eqref{eq:BE0} in terms of the inverse temperature $x \equiv m/T$ and the DM yield $Y \equiv n/s$, where $m$ is the DM mass and $s$ the SM entropy density given by $s(T) \equiv 2 \pi^2/45\,\gss(T)\, T^3$, as
\begin{equation} \label{eq:BE1}
    \frac{dY}{dx} = -\frac{s\, \sv}{x\, H} \left[1 + \frac13\, \frac{d\ln\gss}{d\ln T}\right] \left(Y^2 - Y_{\rm eq}^2\right).
\end{equation}
In the standard cosmological scenario, Eq.~\eqref{eq:BE1} admits two disconnected solutions (and therefore two qualitatively different regimes) to accommodate the entire observed DM relic abundance. If interactions with the SM plasma are efficient enough to establish chemical equilibrium, the DM abundance subsequently freezes out when the interaction rate becomes comparable to the Hubble expansion rate. This typically occurs for masses at the electroweak scale and portal couplings of order $\mathcal{O}(1)$, and corresponds to the WIMP paradigm. Conversely, if chemical equilibrium is never attained, DM can be produced through freeze-in. This typically occurs for very suppressed portal couplings of order $\mathcal{O}(10^{-11})$, and corresponds to the FIMP paradigm.\footnote{The FIMP solution requires small interaction rates, and in the standard cosmological scenario, this can be achieved either by small couplings, as in the case of IR freeze-in, or by high New-Physics scales, as in the case of UV freeze-in~\cite{Elahi:2014fsa}.}

However, in reality, the distinction between these regimes is determined by the efficiency of chemical equilibration, that is, the \textit{reaction rate} rather than solely by the magnitude of the portal couplings. If the thermal production rates become sufficiently Boltzmann suppressed, as can occur in low-$\Trh$ scenarios, chemical equilibrium may never be established despite relatively strong microscopic interactions. It is also interesting to note that, in the case of WIMPs, before its freeze-out the DM density reaches equilibrium. The equilibrium distribution acts as an attractor, erasing the dependence on the initial DM abundance, provided that chemical equilibrium is established sufficiently early. Conversely, in the freeze-in regime, the DM abundance generally retains memory of any pre-existing population and depends on the integrated production history.

In general cosmological scenarios in which the SM entropy is not conserved, it is more convenient to handle Eq.~\eqref{eq:BE0} in terms of the scale factor and the comoving number density $N \equiv n\, a^3$, as
\begin{equation} \label{eq:BE2}
    \frac{dN}{da} = - \frac{\sv}{H\, a^4} \left(N^2 - N_{\rm eq}^2\right),
\end{equation}
where $N_{\rm eq} \equiv n_{\rm eq}\, a^3$. The formulation of Eq.~\eqref{eq:BE2} allows to capture all the features of Eq.~\eqref{eq:BE1} in the case of a standard cosmological scenario, while allowing to explore cases for which the SM entropy is not conserved, as is the case during cosmic reheating. Unlike the yield $Y$, the comoving number density $N$ is not directly affected by entropy production, allowing us to distinguish changes in the particle abundance from entropy dilution during reheating.

Between the conventional freeze-out and freeze-in regimes, intermediate solutions can arise in which DM never reaches chemical equilibrium, although its abundance becomes sufficiently large for annihilation backreactions to be relevant. In this regime, neither the conventional freeze-out approximation nor the freeze-in approximation neglecting backreactions is fully applicable, and the complete Boltzmann equations must be solved, as will be seen in the following. We emphasize that Eq.~\eqref{eq:BE2} is a schematic single-species equation, whereas numerical calculations employ a coupled system.

Finally, to match the entire observed DM relic density, it is required that
\begin{equation}
    m\, Y_0 = \frac{\Omega h^2\, \rho_c}{s_0\, h^2} \simeq 4.3 \times 10^{-10}~\text{GeV},
\end{equation}
where $Y_0$ is the asymptotic value of the DM yield $Y(T)$ at low temperatures after the heavier inert particles have decayed, $s_0 \simeq 2.89 \times 10^3$~cm$^{-3}$ is the present entropy density~\cite{ParticleDataGroup:2024cfk}, $\rho_c \simeq 1.05 \times 10^{-5}~h^2$~GeV/cm$^3$ is the critical energy density of the Universe, and $\Omega h^2 \simeq 0.12$ is the observed DM relic abundance~\cite{Planck:2018vyg}.

%%%%%%%%%%%%%%%%%%%%%%%%%%%%%%%%%%%%%%%%%%%%%%%%%%%%%%%%%%
\section{The Inert Doublet Model} \label{sec:idm}
%%%%%%%%%%%%%%%%%%%%%%%%%%%%%%%%%%%%%%%%%%%%%%%%%%%%%%%%%%
The IDM is an extension of the SM by an exotic $SU(2)_L$ doublet complex scalar field $\Phi$ that is odd under a discrete $\mathbb{Z}_2$ symmetry, under which all SM particles are even. Then, as long as this discrete symmetry remains unbroken, the lightest component of the exotic doublet is stable and can play the role of a DM candidate.

In addition to the gauge interactions arising from its covariant kinetic term, the IDM introduces the following scalar potential
\begin{equation}
    V_0 = \mu_1^2 |H|^2  + \mu_2^2|\Phi|^2 + \lambda_1 |H|^4+ \lambda_2 |\Phi|^4 + \lambda_3 |H|^2| \Phi|^2 + \lambda_4 |H^\dagger\Phi|^2 + \frac{\lambda_5}{2} \Bigl[ (H^\dagger\Phi)^2 + \mathrm{H.c.} \Bigr],
    \label{Eq:TreePotential}
\end{equation}
where $H$ is the usual SM Higgs doublet. For simplicity, in everything that follows, the couplings $\lambda_i$ will be taken to be real. Once the electroweak symmetry is broken, the two doublets can be expanded as
\begin{equation}
	H = \left(
    \begin{array}{c}
        G^+ \\
        \frac{1}{\sqrt{2}}\left(v+h+\mathrm{i}G^0\right)
    \end{array}
    \right), \qquad
	\Phi = \left(
    \begin{array}{c}
        H^+\\
        \frac{1}{\sqrt{2}}\left(H^0+\mathrm{i}A^0\right)
    \end{array}
    \right),
\end{equation}
where $v \simeq 246$~GeV is the usual Higgs vacuum expectation value, $h$ is the physical SM Higgs boson and $G^0/G^{\pm}$ are the neutral and charged Goldstone bosons, respectively. All in all, the IDM introduces four additional scalar degrees of freedom, namely a neutral CP-even scalar $H^0$, a pseudo-scalar $A^0$, and a pair of charged scalars $H^\pm$, along with five new free parameters
\begin{equation} \label{eq:lambdas}
    \left\{\mu_2, ~~\lambda_2, ~~ \lambda_3, ~~ \lambda_4, ~~ \lambda_5 \right\}.
\end{equation}
The tree-level masses of the exotic scalars are related to the parameters of the Lagrangian in Eq.~\eqref{Eq:TreePotential} through
\begin{align}
	m_{H^0}^2 &= \mu_2^2 + \lambda_L\, v^2, \label{Eq:mH0tree} \\
	m_{A^0}^2 &= \mu_2^2 + \lambda_S\, v^2, \\
	m_{H^\pm}^2 &= \mu_2^2 + \frac12\, \lambda_3\, v^2,
\end{align}
where we have conventionally defined
\begin{align}
	\lambda_L &\equiv \frac12 \left( \lambda_3 + \lambda_4 + \lambda_5 \right), \\
	\lambda_S &\equiv \frac12 \left( \lambda_3 + \lambda_4 - \lambda_5 \right).
\end{align}
In practice, $\lambda_{L/S}$ correspond to the couplings of a pair of $H^0$ or $A^0$ particles to the SM Higgs boson, respectively. Instead of the original parameters of Eq.~\eqref{eq:lambdas}, in everything that follows, we will, instead, employ the physically more intuitive ones
\begin{equation}
 	\left\{ m_{H^0}, ~~ m_{A^0}, ~~ m_{H^\pm}, ~~ \lambda_L, ~~ \lambda_2 \right\}.
	\label{eq:masses}
\end{equation}
In addition, and without loss of generality, we choose the mass hierarchy $m_{H^0} < m_{A^0/H^\pm}$, \textit{i.e.} we consider $H^0$ as our DM candidate. The corresponding $A^0$ scenario can be obtained through an appropriate redefinition of the inert-doublet field and its couplings. Moreover, for simplicity, we will consider that $A^0$ and $H^\pm$ are mass-degenerate, with mass $m_{A^0} = m_{H^\pm} = m_{H^0} + \Delta m$. Although this restriction does affect the precise relic density predictions through the coannihilation processes, we do not expect it to qualitatively modify the production mechanisms investigated in this work. In terms of couplings, the equality $m_{A^0} = m_{H^\pm}$ implies $\lambda_4 = \lambda_5$, while $\Delta m > 0$ requires $\lambda_4 = \lambda_5 < 0$. Lastly, the $\lambda_2$ parameter does not directly affect DM production from the SM and is not relevant for the subsequent analysis.

%%%%%%%%%%%%%%%%%%%%%%%%%%%%%%%%%%%%%%%%%%%%%%%%%%%%%%%%%%
\subsection{The IDM as a model of WIMPs} \label{sec:dmidm}
%%%%%%%%%%%%%%%%%%%%%%%%%%%%%%%%%%%%%%%%%%%%%%%%%%%%%%%%%%
The IDM has long been recognized as one of the most representative WIMP DM models: it features a DM candidate, which can be either $H^0$ or $A^0$, that interacts through the electroweak force -- a feature which, in the context of standard cosmology, ensures its early-time thermalization with the SM plasma. Given the model's particle content and Lagrangian, the dark  matter candidate can reproduce the observed DM abundance in the Universe through the three characteristic annihilation mechanisms~\cite{Gondolo:1990dk, Griest:1990kh}: usual annihilation into SM particles, resonant annihilation through the Higgs boson, and/or coannihilation between $H^0$ and $A^0$ or $H^\pm$.

The DM relic density in the IDM has been studied in great detail, \textit{e.g.}, in Refs.~\cite{LopezHonorez:2006gr, LopezHonorez:2010tb, Goudelis:2013uca, Eiteneuer:2017hoh}. In short, the model can saturate the Planck measurement in two broad DM mass ranges, for a recent account, \textit{cf e.g.}~Ref.~\cite{Eiteneuer:2017hoh}: 
\begin{enumerate}
    \item Below $m_{H^0} \simeq 120$~GeV, through annihilations and coannihilations into SM fermions ($m_{H^0} \lesssim 65$~GeV) or (potentially virtual) SM gauge bosons ($65$~GeV $ \lesssim m_{H^0} \lesssim 120$~GeV). Above this mass value, DM turns out to be underabundant due to overly efficient annihilation and coannihilations into SM gauge and Higgs boson pairs.
    \item However, for $m_{H^0} \gtrsim 500$~GeV, destructive interference between Feynman diagrams can become efficient enough so that the DM density constraint can be satisfied again. In this regime, the predicted relic density depends on: $i)$ the mass difference between $H^0$ and the heavier states; within the parameter range considered here, increasing the mass splitting generally enhances the effective depletion rate of the inert sector, leading to a smaller relic abundance. And $ii)$ on the value of $\lambda_L$; negative values of $\lambda_L$ being partly capable of counter-balancing a larger mass splitting but, at the same time, giving rise to annihilations into Higgs boson pairs.
\end{enumerate}
With the exception of the Higgs resonance $m_{H^0} \simeq m_h/2 \simeq 62.5$~GeV, the IDM low-mass region is essentially excluded by direct detection experiments~\cite{Eiteneuer:2017hoh}, at least in the context of standard cosmology. One possibility to re-open part of the $m_{H^0} \lesssim 500$~GeV mass range was studied in Refs.~\cite{Barman:2021ifu, Belanger:2025ack}, in the context of ``fast expansion'' cosmologies like kination. In this case, scenarios in which DM would otherwise be underabundant become viable because a faster cosmic expansion leads to earlier freeze-out. In this paper, we are rather interested in regions of parameter space in which DM tends to be \textit{overabundant}. For this reason, in everything that follows, we will focus on the high-mass region of the parameter space, $m_{H^0} \gtrsim 500$~GeV.

%%%%%%%%%%%%%%%%%%%%%%%%%%%%%%%%%%%%%%%%%%%%%%%%%%%%%%%%%%
\subsection{Constraints on the IDM parameter space} \label{sec:constraints}
%%%%%%%%%%%%%%%%%%%%%%%%%%%%%%%%%%%%%%%%%%%%%%%%%%%%%%%%%%
In general, the IDM is subject to a wide variety of bounds. However, in our mass range of interest, there are four main sources of constraints.
\begin{itemize}
    \item First, constraints from direct searches for WIMP scattering off nuclei. These impose a direct bound on the absolute value of $\lambda_L$, since the $H^0$-quark scattering cross-section is proportional to $\lambda_L^2$. In our analysis we require the tree-level WIMP-nucleon scattering cross-section to be smaller than the latest limits from LZ~\cite{LZ:2024zvo}. We find that for masses $m_{H^0} \lesssim 1$~TeV, this leads to a restriction $|\lambda_L| \lesssim 0.09$, with the precise bound becoming stronger for lower masses and reaching $|\lambda_L| \lesssim 0.03$ for $m_{H^0} \simeq 500$~GeV, \textit{cf} also Ref.~\cite{Belanger:2025ack}. Future direct-detection experiments may probe parts of the IDM high-mass region even in the limit $\lambda_L = 0$, due to electroweak loop contributions~\cite{Klasen:2013btp}.
    \item These limits on the value $\lambda_L$ make it so that Higgs-mediated DM annihilation can only be moderate. The relic density is driven, rather, by a combination of annihilations and coannihilations into SM gauge bosons which is mainly determined by the mass splitting between $H^0$ and the heavier scalars, with additional tuning that depends on the (necessarily small) value of $\lambda_L$. The mass splitting can be constrained from Fermi-LAT searches for DM annihilation-induced gamma-rays in dwarf spheroidal galaxies (dSphs)~\cite{Fermi-LAT:2013sme, Calore:2018sdx, Alvarez:2020cmw}. The recent analysis performed in Ref.~\cite{Belanger:2025ack} showed that, in the mass range $500 \ {\rm{GeV}} < m_{H^0} < 1 \ {\rm{TeV}}$, $\Delta m$ cannot exceed $60$~GeV, with the bound becoming as strong as $\Delta m \lesssim 35$~GeV for $m_{H^0} \simeq 500$~GeV. The precise upper bound on $\Delta m$ depends on the DM mass and the remaining IDM parameters, in particular $\lambda_L$, as discussed in Ref.~\cite{Belanger:2025ack}.
    \item In the limit in which $m_{H^0}$ approaches sufficiently the mass of $H^\pm$, the latter can become long-lived. Depending on the precise value of $\Delta m$, its decay width can be dominated \textit{i}) either by the 3-body decay $H^\pm \to H^0\ell^\pm\nu_\ell$, which occurs for $\Delta m \gtrsim 500$~MeV and for $\Delta m < m_{\pi^\pm}\simeq 140$~MeV,  or \textit{ii}) by the two-body decay $H^\pm \to H^0\pi^\pm$, when $140~{\rm MeV}\lesssim\Delta_m\lesssim 500~{\rm MeV}$. A lower bound on the $H^0$ - $H^\pm$ mass difference can be obtained from LHC searches for Heavy Stable Charged Particles (HSCPs) \cite{ATLAS:2019gqq, CMS:2015lsu}, which we have quantified using {\tt SModelS}~v3~\cite{Altakach:2024jwk} through its interface with {\tt micrOMEGAs~7}~\cite{Belanger:2026asz}. The pion decay mode has been implemented through an effective Lagrangian along the lines described in Ref.~\cite{Belyaev:2016lok}. Our numerical findings show that these searches can impose a limit $\Delta m > 200$~MeV for $m_{H^0}\lesssim 450$~GeV, which decreases to $\Delta m > 100$~MeV for masses $m_{H^0}\simeq 500$~GeV. Above roughly 530~GeV, no meaningful constraint can be obtained. We also note that searches for disappearing tracks do not impose any limit in our parameter space of interest.
    \item Lastly, we impose the theoretical consistency requirements of boundedness from below, perturbative unitarity, and perturbativity using the {\tt 2HDMC} package~\cite{Eriksson:2009ws}. We also require the inert vacuum to be the global minimum of the scalar potential. Within the parameter space considered, and for the adopted treatment of $\lambda_2$, these constraints do not lead to additional exclusions beyond those imposed by the experimental bounds.
\end{itemize}
The experimental bounds discussed above are independent of the pre-BBN cosmological history, provided that the present-day DM distribution is unchanged and $H^0$ accounts for the entire observed DM abundance.

For our numerical analysis, we will employ the {\tt micrOMEGAs~7}~\cite{Belanger:2001fz, Alguero:2023zol, Belanger:2026asz} package to compute all necessary DM observables. In particular, it will be used for the simultaneous resolution of the background Eqs.~\eqref{eq:background1} to~\eqref{eq:background3} along with the necessary Boltzmann equations for all states out-of-equilibrium as described in Refs.~\cite{Alguero:2023zol, Belanger:2026asz}. We evolve the inert species separately and include their decays and conversions. The initial DM abundance (together with the abundance of all other dark sector particles) is consistently set to zero, which will allow us to capture both freeze-out and freeze-in - like behaviors. On the side of cosmology, in the instantaneous reheating approximation the only free parameter is $\Trh$. In the finite-duration reheating case, reheating is parametrized by $H_I$ and $\Gamma_\phi$. However, in practice, throughout our analysis, we will argue in terms of the more physically intuitive parameters $\Trh$ and $\Tmax$.

%%%%%%%%%%%%%%%%%%%%%%%%%%%%%%%%%%%%%%%%%%%%%%%%%%%%%%%%%%
\subsection{The IDM relic density in standard cosmology} \label{sec:IDMstandard}
%%%%%%%%%%%%%%%%%%%%%%%%%%%%%%%%%%%%%%%%%%%%%%%%%%%%%%%%%%
As we already mentioned in Section~\ref{sec:dmidm}, the IDM relic density within standard cosmology is determined, for a given DM mass, by two parameters: the mass difference between DM and the heavier states, $\Delta m$, and the value of $\lambda_L$. At the same time, as described in Section~\ref{sec:constraints}, the constraints arising from direct DM searches impose strong restrictions on the latter. Once these constraints are imposed, the range of $\Delta m$ for which the IDM can account for the entire observed DM abundance in the Universe also turns out to be fairly narrow for each value of $m_{H^0}$.

\begin{figure}[t!]
    \def\sepf{0.496}
    \centering
    \includegraphics[width=0.8\columnwidth]{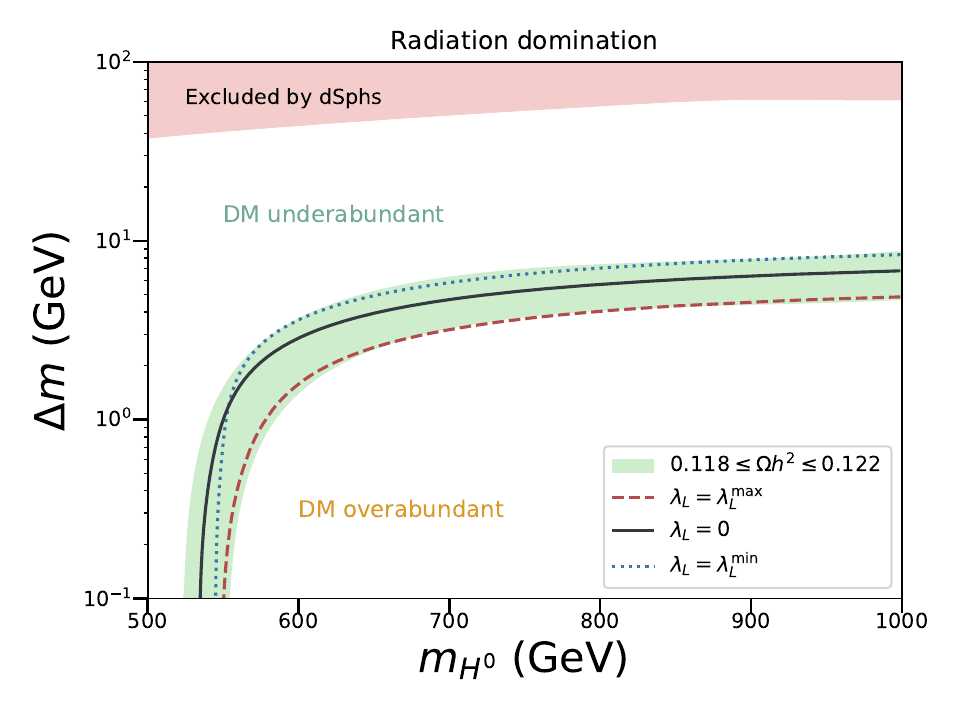}
    \caption{The cosmologically viable parameter space of the IDM projected on the $(m_{H^0}, \Delta m)$ plane, assuming a standard thermal history (green shaded area). Also shown are contours of $\Omega h^2 = 0.12$ for three benchmarks, namely $\lambda_L = 0$ (black solid line) and $\lambda_L$ fixed, for each DM mass, to its maximal or minimal value allowed by direct detection constraints (dashed red and dotted blue lines, respectively). The red shaded region is excluded by indirect searches for gamma-rays in dSphs. HSCP limits, which apply for masses $m_{H^0} \lesssim 530$~GeV and lie below $\Delta m = 100$~MeV are not shown.}
    \label{fig:standardresult}
\end{figure}

In Fig.~\ref{fig:standardresult} we show the IDM cosmologically viable parameter space  (corresponding to $0.118< \Omega h^2 < 0.122$) in the $(m_{H^0}, \Delta m)$ plane in the standard cosmological scenario, once the coupling $\lambda_L$ is varied within its entire experimentally allowed range (shaded green area). We, moreover, highlight three distinct benchmarks: the case $\lambda_L = 0$ (solid black line), in which the DM-nucleon scattering cross-section vanishes at tree-level, as well as the cases in which $\lambda_L$ is fixed to its maximal (positive) or minimal (negative) value allowed by direct detection (dashed-red and dotted-blue lines, respectively). Constraints from Fermi-LAT are depicted by the red shaded area.
%%%%%%%%%%%%%%%%%%%%%%%%%%%%%%%%%%%%%%%%%%%%%%%%%%%%%%%%%%
%%%%%%%%%%%%%%%%%%%%%%%%%%%%%%%%%%%%%%%%%%%%%%%%%%%%%%%%%%

As we observe, the minimal and maximal $\lambda_L$ scenarios bracket the viable parameter space, with the exception of the lowest allowed $H^0$ masses. The relic density is obtained through a non-trivial combination of annihilations and coannihilations which varies according to the mass splitting and the sign of $\lambda_L$. The viable mass splitting values turn out to be even more restricted than allowed by dSphs constraints, with $\Delta m$ bounded to be smaller than $10$~GeV for $m_{H^0} \lesssim 1$~TeV. For larger mass splittings, DM turns out to be underabundant. As we already mentioned, these regions of parameter space can again become viable, \textit{e.g.}, in kination scenarios~\cite{Belanger:2025ack} and can be constrained by indirect DM searches. Smaller mass splittings, on the other hand, lead to the Universe being overclosed. These regions of parameter space will be of particular interest in the following.

%%%%%%%%%%%%%%%%%%%%%%%%%%%%%%%%%%%%%%%%%%%%%%%%%%%%%%%%%%
\section{The IDM relic density in scenarios of low-temperature reheating} \label{sec:results}
%%%%%%%%%%%%%%%%%%%%%%%%%%%%%%%%%%%%%%%%%%%%%%%%%%%%%%%%%%
The previous picture becomes substantially modified once we move beyond the standard cosmological assumptions of high-temperature reheating. In the following paragraphs, we revisit the IDM relic density predictions in two alternative scenarios: first, assuming that the Universe reheated instantaneously to a sufficiently low temperature. Subsequently, we consider the case of finite-duration reheating.

%%%%%%%%%%%%%%%%%%%%%%%%%%%%%%%%%%%%%%%%%%%%%%%%%%%%%%%%%%
\subsection{The relic density assuming instantaneous reheating} \label{sec:instantaneous}
%%%%%%%%%%%%%%%%%%%%%%%%%%%%%%%%%%%%%%%%%%%%%%%%%%%%%%%%%%
%%%%%%%%%%%%%%%%%%%%%%%%%%%%%%%%%%%%%%%%%%%%%%%%%%%%%%%%%%
\begin{figure}[t!]
    \def\sepf{0.496}
    \centering
    \includegraphics[width=0.8\columnwidth]{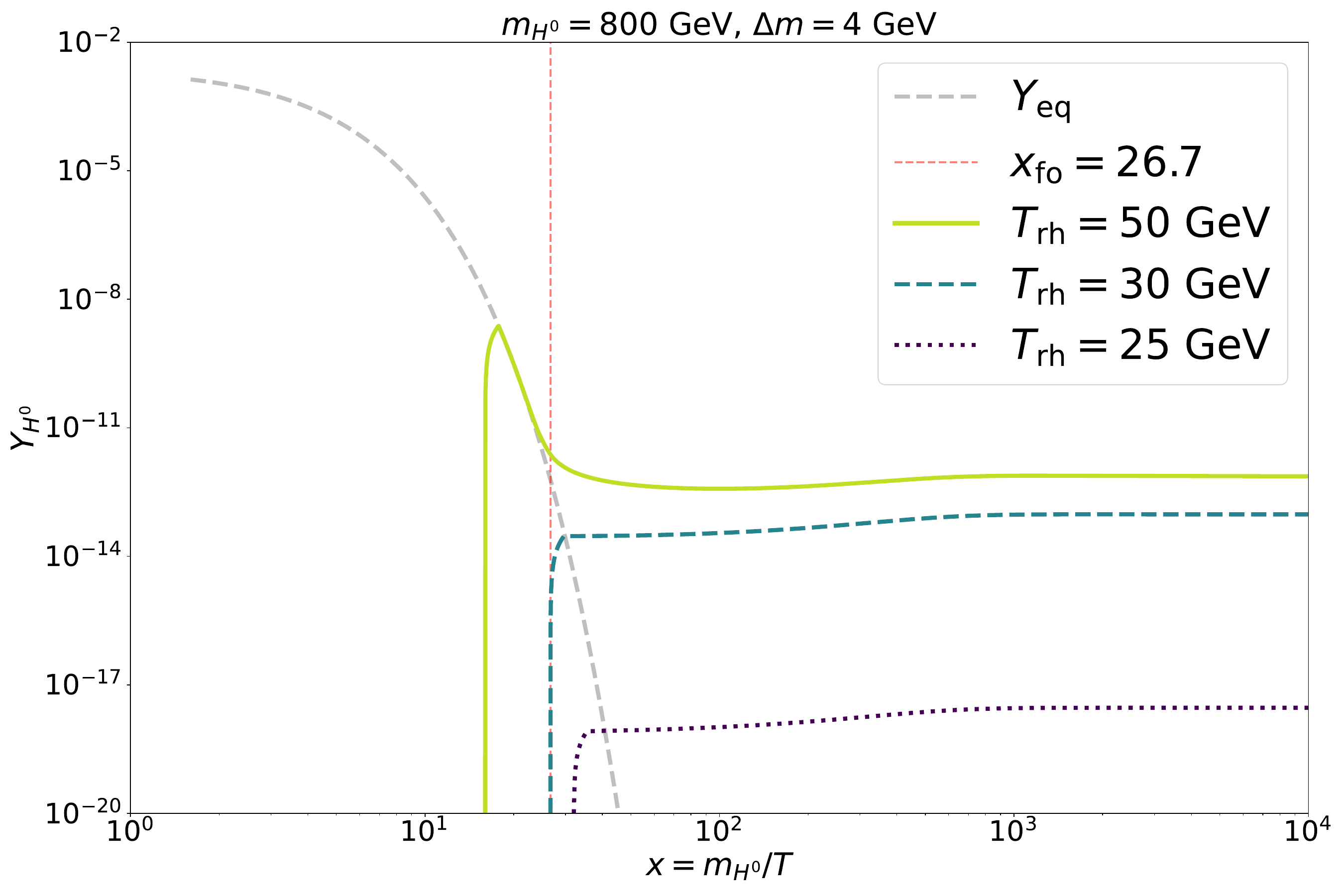}
    \caption{DM yield evolution as a function of $x \equiv m_{H^0}/T$ for a benchmark with $m_{H^0} = 800$~GeV, $\lambda_L = 0$ and $\Delta m = 4$~GeV, for $\Trh = 50$, $30$ and $25$~GeV (solid green, dashed blue and dotted purple curves, respectively). The equilibrium density is represented by the dashed gray curve whereas $x_{\rm fo}$ by the vertical, dotted red line.}
    \label{fig:yiedlTrh}
\end{figure} 
%%%%%%%%%%%%%%%%%%%%%%%%%%%%%%%%%%%%%%%%%%%%%%%%%%%%%%%%%%
Let us first start with the simplest scenario of instantaneous reheating. In Fig.~\ref{fig:yiedlTrh} we show the evolution of the DM abundance as a function of $x \equiv m_{H^0}/T$, for a benchmark with $m_{H^0} = 800$~GeV, $\lambda_L = 0$ and $\Delta m = 4$~GeV, assuming three different $\Trh$ values, namely $\Trh = 50$, $30$ and $25$~GeV (solid green, dashed blue and dotted purple curves, respectively). The dashed gray line depicts the equilibrium abundance, whereas the dotted red vertical line corresponds to $x_{\rm fo}$, the $x$ value for which $H^0$ would freeze-out under standard cosmology. In this scenario, it corresponds to $x_{\rm fo}\simeq 26.7$. The DM relic density in the three cases is found to be $\Omega h^2 = 0.16$, $0.02$, and $6.4 \times 10^{-7}$, respectively. 

As we observe, as long as the reheating temperature is sufficiently above the would-be freeze-out temperature, interactions with the SM plasma can establish chemical equilibrium, and the DM abundance subsequently evolves according to the conventional thermal freeze-out picture presented in Section~\ref{sec:IDMstandard}. Concretely, even if we assume that DM was initially absent from the thermal bath, its abundance increases rapidly until it reaches its equilibrium value and, subsequently, freezes out. 

As the reheating temperature approaches or falls below the would-be freeze-out temperature, chemical equilibration becomes increasingly inefficient. Eventually, DM never attains equilibrium and, assuming again that its initial density is zero, it follows a freeze-in like evolution, rapidly settling to an abundance which is suppressed for lower and lower values of $\Trh$.\footnote{Here, again, $\Tfo$ is to be understood as the \textit{would-be} freeze-out temperature: the temperature at which DM would freeze out in the standard cosmological scenario. In the freeze-in regime this equilibrium is never reached, but $\Tfo$ remains a useful reference scale for comparing against $\Trh$.} Figure~\ref{fig:yiedlTrh} illustrates exactly this fact: once the fraction of the plasma which can produce DM becomes sufficiently Boltzmann-suppressed, early-time DM-SM equilibration is no longer guaranteed and conventional WIMP-type DM candidates exhibit a freeze-in behavior. We note that the selected benchmark is illustrative; the three reheating temperatures were not chosen to reproduce the observed relic abundance, but to show the impact of lowering $\Trh$.

%%%%%%%%%%%%%%%%%%%%%%%%%%%%%%%%%%%%%%%%%%%%%%%%
\begin{figure}[t!]
    \def\sepf{0.496}
    \centering
    \includegraphics[width=0.8\columnwidth]{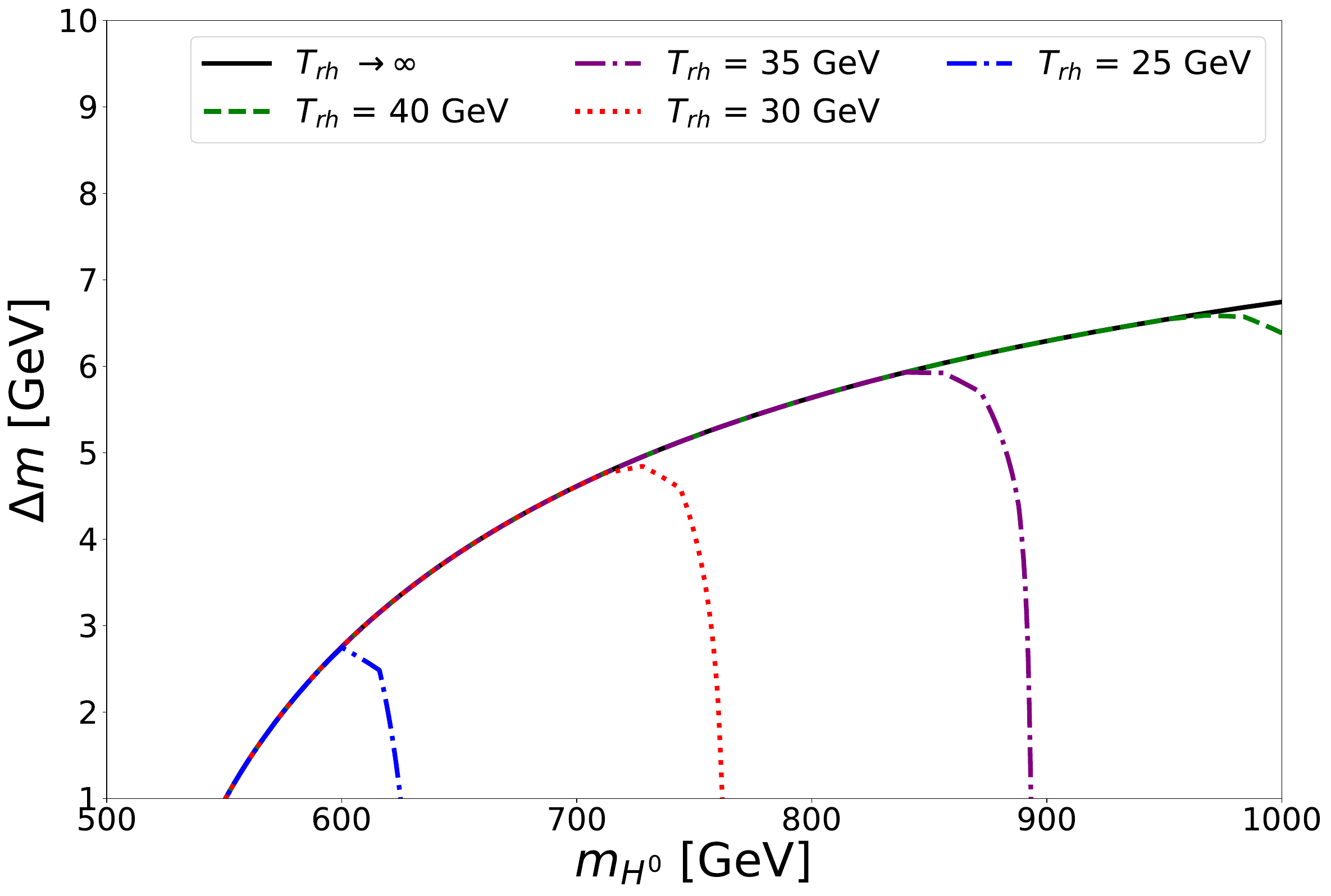}
    \caption{Contours that satisfy the relic density constraint for different values of the (instantaneous) reheating temperature, for $\lambda_L = 0$.}
    \label{fig:LowTrh}
\end{figure} 
%%%%%%%%%%%%%%%%%%%%%%%%%%%%%%%%%%%%%%%%%%%%%%%%
As an additional remark, we note that, at late times, the DM yield continues to increase until $x \sim \mathcal O(10^3)$, before approaching its asymptotic value. This behavior is caused by conversions and decays of the heavier inert states into the stable DM candidate, which primarily redistribute the existing inert-sector population rather than generate additional inert particles from the SM plasma, as further explained in Appendix~\ref{sec:lowT}. We emphasize that the late-time increase of the $H^0$ yield does not imply continued production of inert particles from the SM bath, but rather reflects the redistribution of the previously produced inert-sector population.

Let us now turn to the overall impact of the previous considerations on the cosmologically viable parameter space of the IDM. For simplicity, we first consider the most canonical WIMP case $\lambda_L = 0$, for which electroweak gauge interactions remain present. In Fig.~\ref{fig:LowTrh} we show the contours of $\Omega h^2 = 0.12$ in the $(m_{H^0}, \Delta m)$ plane for different values of the reheating temperature, $\Trh \gg 40$~GeV (in practice this corresponds to  standard cosmology since, for the parameters considered, the DM production occurs at lower temperatures), $\Trh =40$, 35, 30 and 25~GeV (solid black, dashed green, dot-dashed violet, dotted red, and dot-dashed blue lines, respectively). For a given value of $\Trh$ the $(m_{H^0}, \Delta m)$ values for which the IDM can explain DM fall into two categories: \textit{i}) one which is identical with the ``effectively infinite'' reheating temperature scenario, \textit{i.e.} the part of the curves following the $\Trh \gg 40$~GeV line, and \textit{ii}) the parts of the curves branching down to lower values of $\Delta m$, \textit{i.e.} into the region in which DM would be overabundant in the standard cosmological scenario. In the first case, the DM mass is sufficiently low such that $\Tfo < \Trh$ and $H^0$ undergoes conventional thermal freeze-out. In the latter case, DM never equilibrates and freezes-in. The branches extending toward smaller mass splittings predominantly correspond to nonequilibrium production. The transition from conventional freeze-out to freeze-in is continuous and may involve intermediate solutions in which the chemical equilibrium is not fully established, although annihilation backreactions remain important.

We emphasize that, generically, a decrease of the reheating temperature translates into a smaller DM yield which, in turn, has to be compensated by an increase of the DM production rate. This increase could come from enhancing generic portal couplings, as in the ``freeze-in at stronger coupling'' scenario~\cite{Cosme:2023xpa, Silva-Malpartida:2023yks}. Here, instead, we have an example in which the portal couplings are \textit{fixed}, since they correspond to the SM gauge interactions. In the $\lambda_L = 0$ IDM benchmark, the only parameters that can be varied are the mass splitting and, eventually, the reheating temperature. We emphasize that, in the IDM, the variation of the mass splitting modifies both the production thresholds of the heavier inert states and the interaction amplitudes entering the different production channels. In particular, reducing $\Delta m$ can enhance the total thermal production of the inert sector, thereby compensating for the suppression associated with a lower reheating temperature. This allows the observed relic abundance to be recovered without modifying the electroweak gauge couplings.

%%%%%%%%%%%%%%%%%%%%%%%%%%%%%%%%%%%%%%%%%%%%%%%%%%%%%%%%%%
\begin{figure}[t!]
    \def\sepf{0.496}
    \centering
    \includegraphics[width=0.8\columnwidth]{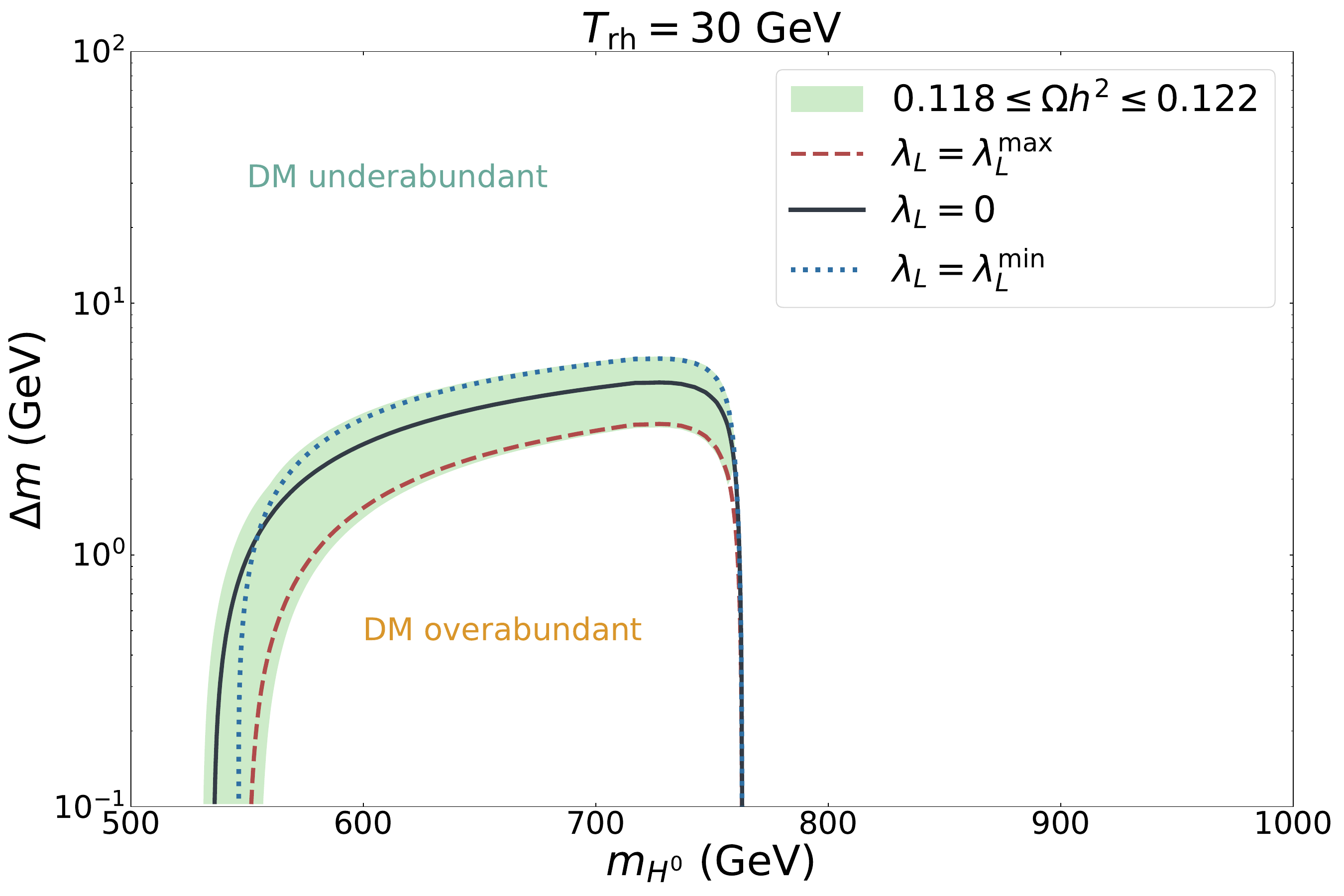}
    \caption{The cosmologically viable parameter space of the IDM projected on the $(m_{H^0}, \Delta m)$ plane, assuming instantaneous reheating at a temperature $\Trh = 30$~GeV (green shaded area). Also shown are contours of $\Omega h^2 = 0.12$ for three benchmarks, namely $\lambda_L = 0$ (black solid line) and $\lambda_L$ fixed, for each DM mass, to its maximal or minimal value allowed by direct detection constraints (dashed red and dotted blue lines, respectively).}
    \label{fig:LowTrhband}
\end{figure} 
%%%%%%%%%%%%%%%%%%%%%%%%%%%%%%%%%%%%%%%%%%%%%%%%%%%%%%%%%%
Lastly, to obtain a more complete picture of how the cosmologically viable parameter space of the IDM is modified, in Fig.~\ref{fig:LowTrhband} we show the region satisfying the observed relic-density constraint, assuming instantaneous reheating at $\Trh = 30$~GeV and allowing $\lambda_L$ to vary within the range permitted by direct detection.

As we can observe, for relatively low DM masses, the viable parameter space closely follows the standard cosmological result shown in Fig.~\ref{fig:standardresult}. In this region, the reheating temperature is sufficiently high for the inert states to attain chemical equilibrium with the SM plasma, and the relic abundance is determined by conventional thermal freeze-out. However, as the DM mass increases, the corresponding freeze-out temperature approaches $\Trh$, and the finite initial temperature of the radiation bath becomes increasingly important. Eventually, chemical equilibrium can no longer be established, and DM is produced through freeze-in. The transition between these two regimes is continuous, as already illustrated in Fig.~\ref{fig:LowTrh}.

Once again, the viable parameter space extends into regions that would lead to an excessive DM abundance in standard cosmology. In the freeze-in regime, the final abundance is controlled by the integrated production rate rather than by the efficiency of thermal depletion. Consequently, parameter configurations that would yield insufficient annihilation in the conventional freeze-out scenario can reproduce the observed relic abundance once the thermal production of inert particles is suppressed by the low reheating temperature. This behavior is not restricted to the benchmark $\lambda_L = 0$, but persists when the Higgs-portal coupling is varied within its entire experimentally allowed range.

Another noteworthy feature of Fig.~\ref{fig:LowTrhband} is the existence of a maximal DM mass beyond which the observed relic abundance can no longer be reproduced. For $\Trh = 30$~GeV, this occurs at approximately $m_{H^0} \sim 760$~GeV, given the parameter space and experimental constraints considered here. At larger masses, the production of inert particles from the SM thermal bath becomes increasingly Boltzmann suppressed, and the available interactions are no longer sufficient to generate the observed abundance, even after varying $\lambda_L$ and $\Delta m$ within the parameter space considered. This illustrates that a low reheating temperature not only opens previously excluded regions of WIMP parameter space but can also impose an upper limit on the DM mass compatible with the observed relic abundance.

%%%%%%%%%%%%%%%%%%%%%%%%%%%%%%%%%%%%%%%%%%%%%%%%%%%%%%%%%%
\subsection{Beyond the instantaneous reheating approximation} \label{sec:noninstantaneous}
%%%%%%%%%%%%%%%%%%%%%%%%%%%%%%%%%%%%%%%%%%%%%%%%%%%%%%%%%%
As we just saw, once the reheating temperature becomes sufficiently low, an alternative possibility to thermal freeze-out appears in WIMP models: that of freeze-in. Despite their different physical origins, freeze-out and freeze-in share an important feature in the instantaneous-reheating approximation: both occur entirely during the radiation-dominated era.

Despite its simplicity, the instantaneous inflaton decay approximation is generally not believed to provide a realistic picture of the reheating process. As described in Section~\ref{sec:reheatingdynamics}, a finite reheating period involves the gradual creation of the SM thermal bath during an epoch characterized by a modified expansion rate with respect to conventional radiation domination. This leads to the appearance of an additional relevant temperature, namely the maximal temperature $\Tmax$ attained by the plasma. In turn, this means that even if the inflaton decayed exclusively into SM particles, DM production from the SM thermal bath may also have occurred \textit{during} reheating, either through freeze-in or through freeze-out.
 
\begin{figure}[t!]
    \def\sepf{0.496}
    \centering
    \includegraphics[width=0.49\columnwidth]{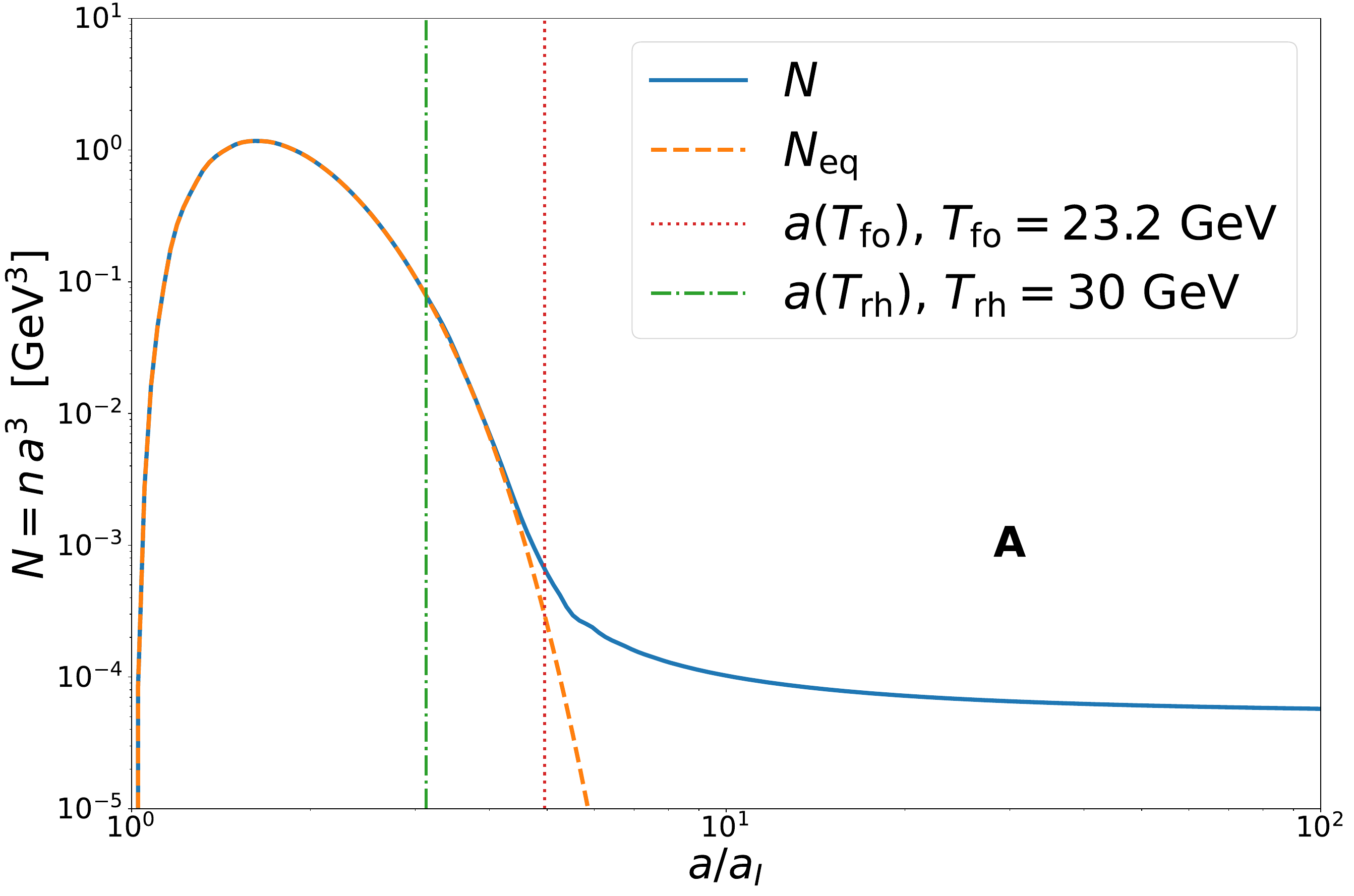} \includegraphics[width=0.49\columnwidth]{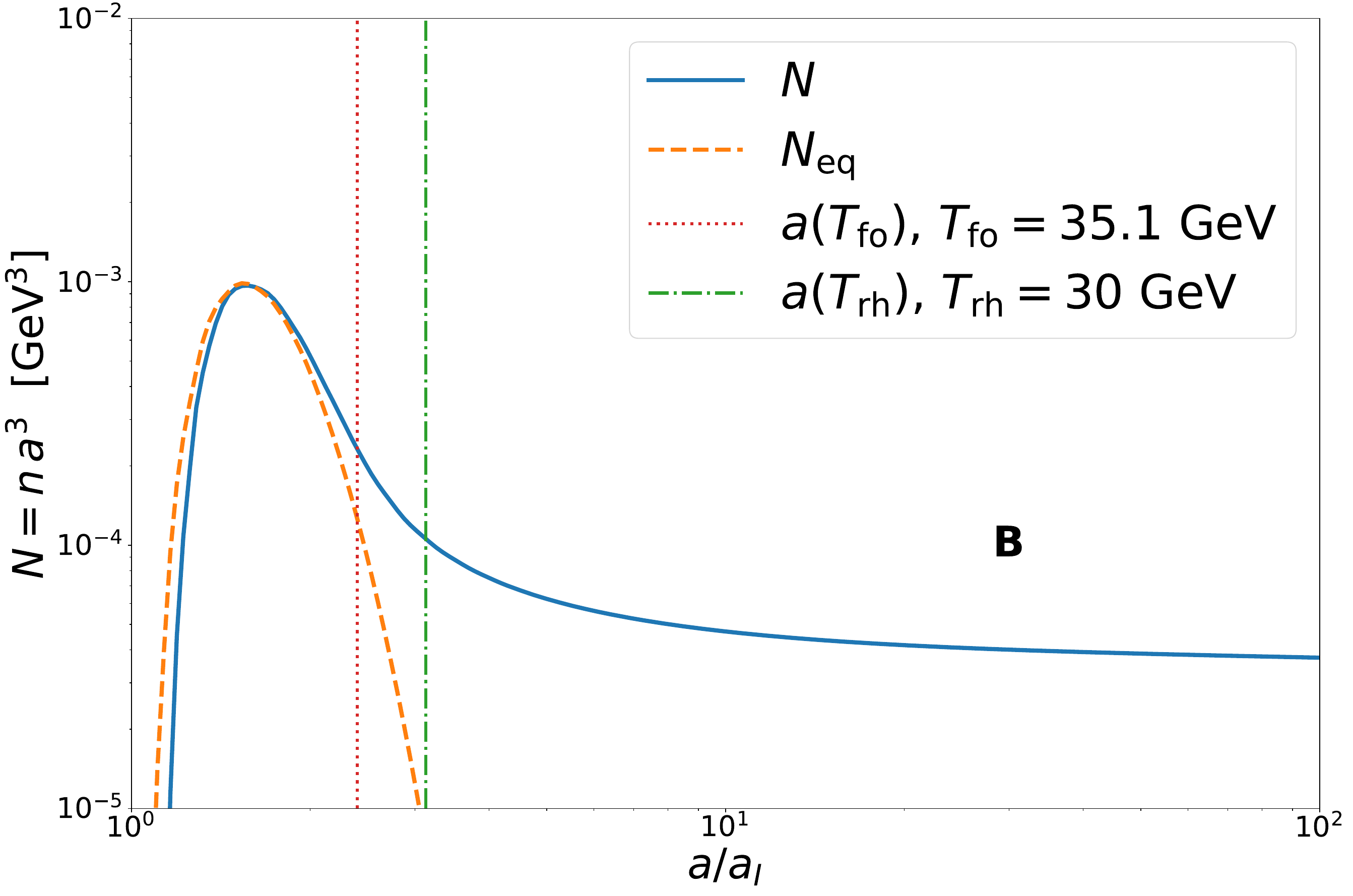} \\
    \includegraphics[width=0.49\columnwidth]{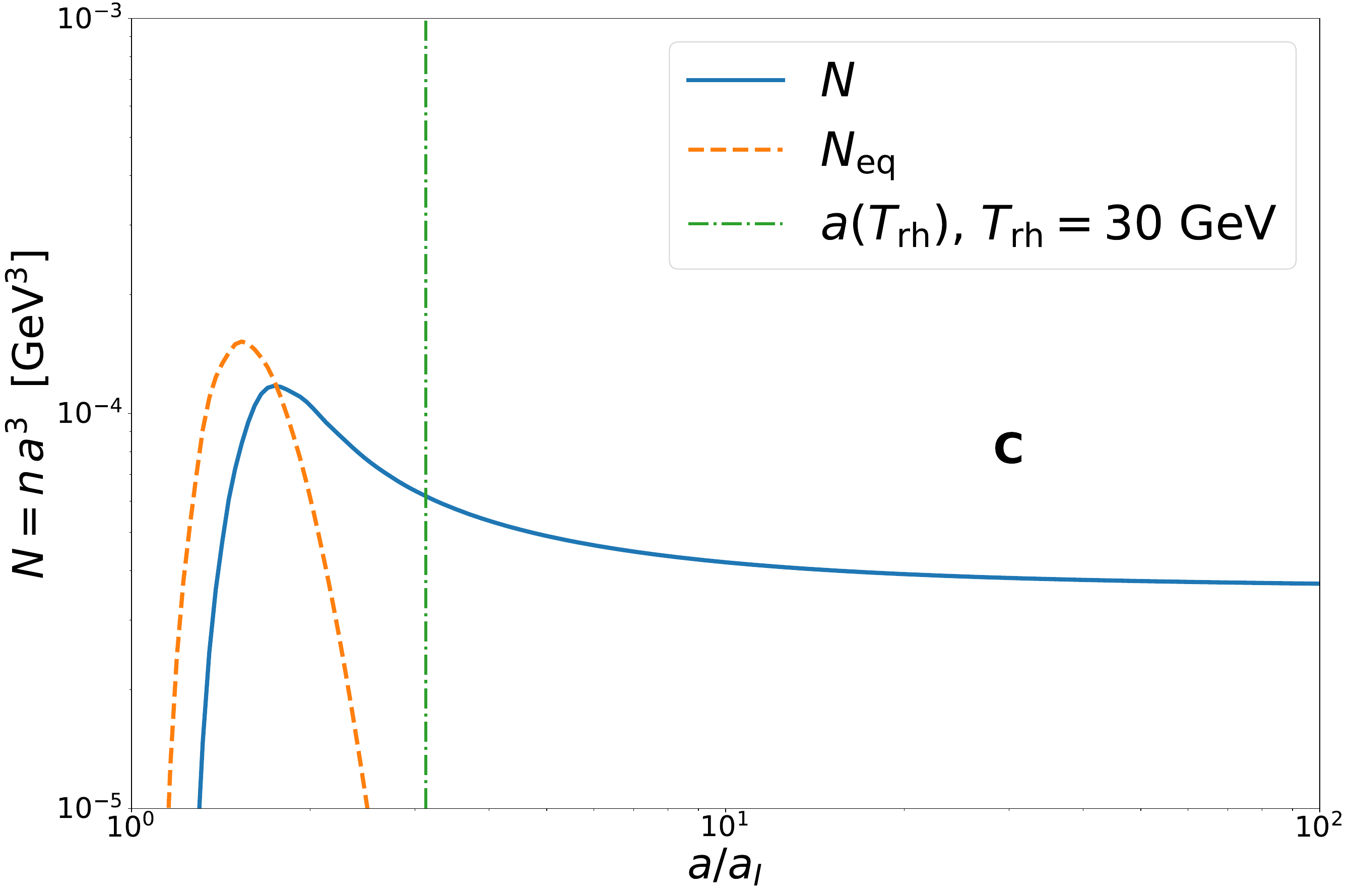} \includegraphics[width=0.49\columnwidth]{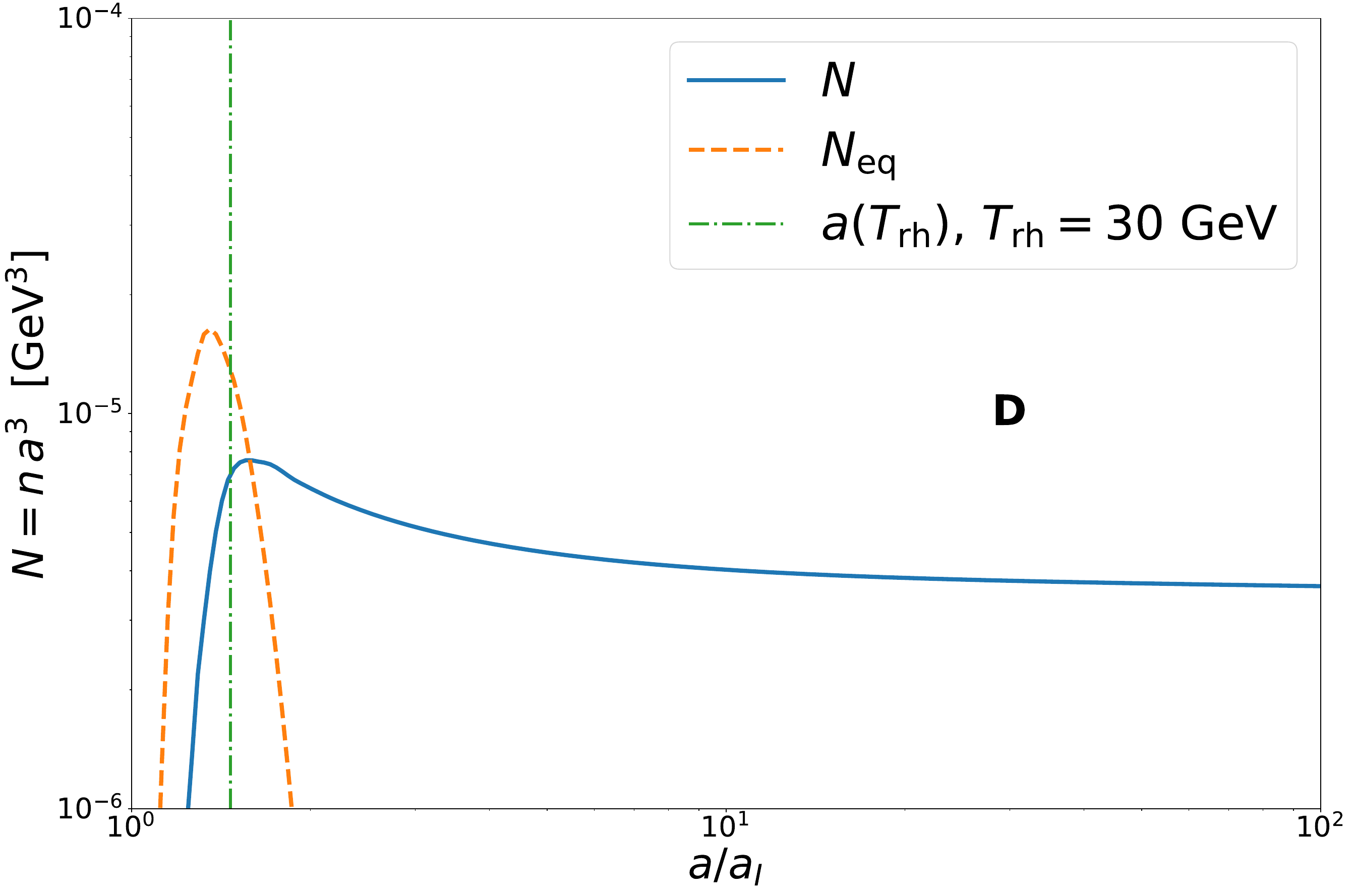}    
    \caption{The DM density evolution (solid blue line) and equilibrium density (dashed orange line) as a function of the scale factor for $m_{H^{0}}=600$~GeV, $\Delta m=2$~GeV, $\Trh=30$~GeV, $\Tmax=40$~GeV (top-left panel). $m_{H^{0}}=904$~GeV, $\Delta m=3.86$~GeV, $\Trh=30$~GeV, $\Tmax=40$~GeV (top-right panel). $m_{H^{0}}=985$~GeV, $\Delta m=2$~GeV, $\Trh=30$~GeV, $\Tmax=40$~GeV (bottom-left panel). $m_{H^{0}}=800$~GeV, $\Delta m=2$~GeV, $\Trh=30$~GeV, $\Tmax=31$~GeV (bottom-right panel).}
    \label{fig:finitereheatingbenchmarks}
\end{figure}

In order to illustrate this feature, in Fig.~\ref{fig:finitereheatingbenchmarks} we show the evolution of $N \equiv n\, a^3$ as a function of the scale factor ratio $a/a_I$ (solid blue lines), for four different benchmarks denoted as A, B, C and D.\footnote{Here $n$ corresponds to the sum of all inert-sector number densities: $n = n_{H^0} + n_{A^0} + n_{H^+} + n_{H^-}$.} With a bit of hindsight, we choose them as $(m_{H^0},\Delta m) = (600,2)$~GeV, $(904,3.86)$~GeV, $(985,2)$~GeV and $(800,2)$~GeV (top-left, top-right, bottom-left and bottom-right panels, respectively). In all four cases, the dashed orange curves depict the corresponding $N_{\rm eq}$ evolution, whereas the vertical dot-dashed green and dotted red lines highlight the scale factor values corresponding to $T = \Trh$ and $T = T_{\rm fo}$, respectively (the corresponding temperatures are quoted in each panel's label). All four scenarios satisfy the relic density constraint for appropriate values of the underlying cosmological parameters, namely $(\Trh, \Tmax) = (30, 40)$~GeV (top-left, top-right and bottom-left panels) and $(30,31)$~GeV (bottom-right panel).
%%%%%%%%%%%%%%%%%%%%%%%%%%%%%%%%%%%%%%%%%%%%%%%%%%%%%%%%%%
%%%%%%%%%%%%%%%%%%%%%%%%%%%%%%%%%%%%%%%%%%%%%%%%%%%%%%%%%%

These benchmarks illustrate the different situations that can arise when reheating is assumed to have a finite duration: freeze-out during radiation domination (Scenario A); freeze-out during reheating (Scenario B). In this case, the final relic abundance is affected not only by the modified expansion rate during chemical decoupling but also by residual annihilations and the entropy injected by subsequent inflaton decays. In addition, we can also have DM freeze-in, with the DM abundance peaking deep during reheating (Scenario C) or peaking approximately around the onset of radiation domination (Scenario D). In this sense, scenarios C and D can be regarded as UV-dominated freeze-in, although the temperature sensitivity arises from Boltzmann suppression rather than from higher-dimensional interactions. Furthermore, Scenario D is the closest analog to the instantaneous reheating scenario. As $\Tmax$ approaches $\Trh$, the viable parameter space increasingly resembles that obtained in the instantaneous-reheating approximation. Nevertheless, the finite-duration scenarios considered here need not reproduce the instantaneous results exactly, since the radiation–inflaton equality temperature does not necessarily coincide with the initial temperature of an instantaneous radiation-dominated cosmology. A precise comparison therefore requires matching the corresponding thermal histories and temperature conventions. In the finite-duration reheating scenario, residual inflaton decays continue to inject energy and entropy into the SM plasma after radiation–inflaton equality, modifying the temperature–scale-factor relation and the subsequent evolution of the DM abundance. For this reason, we will refrain from directly comparing our numerical results with those obtained in the instantaneous reheating approximation.

A further noteworthy observation concerns the two freeze-in scenarios shown in the bottom panels: in both cases, some degree of DM backreaction, through annihilations into SM particles, is present. As pointed out in Refs.~\cite{Cosme:2023xpa, Goudelis:2026lyy}, these backreactions become relevant at temperatures somewhat lower than those at which freeze-in production takes place. In addition, we emphasize that the absence of chemical equilibrium with the SM plasma does not preclude efficient conversions among the inert states, which may maintain relative chemical equilibrium within the inert sector, as discussed in Appendix~\ref{sec:lowT}.
%%%%%%%%%%%%%%%%%%%%%%%%%%%%%%%%%%%%%%%%%%%%%%%%%%%%%%%%%%
\begin{figure}[t!]
    \def\sepf{0.496}
    \centering
    \includegraphics[width=0.8\columnwidth]{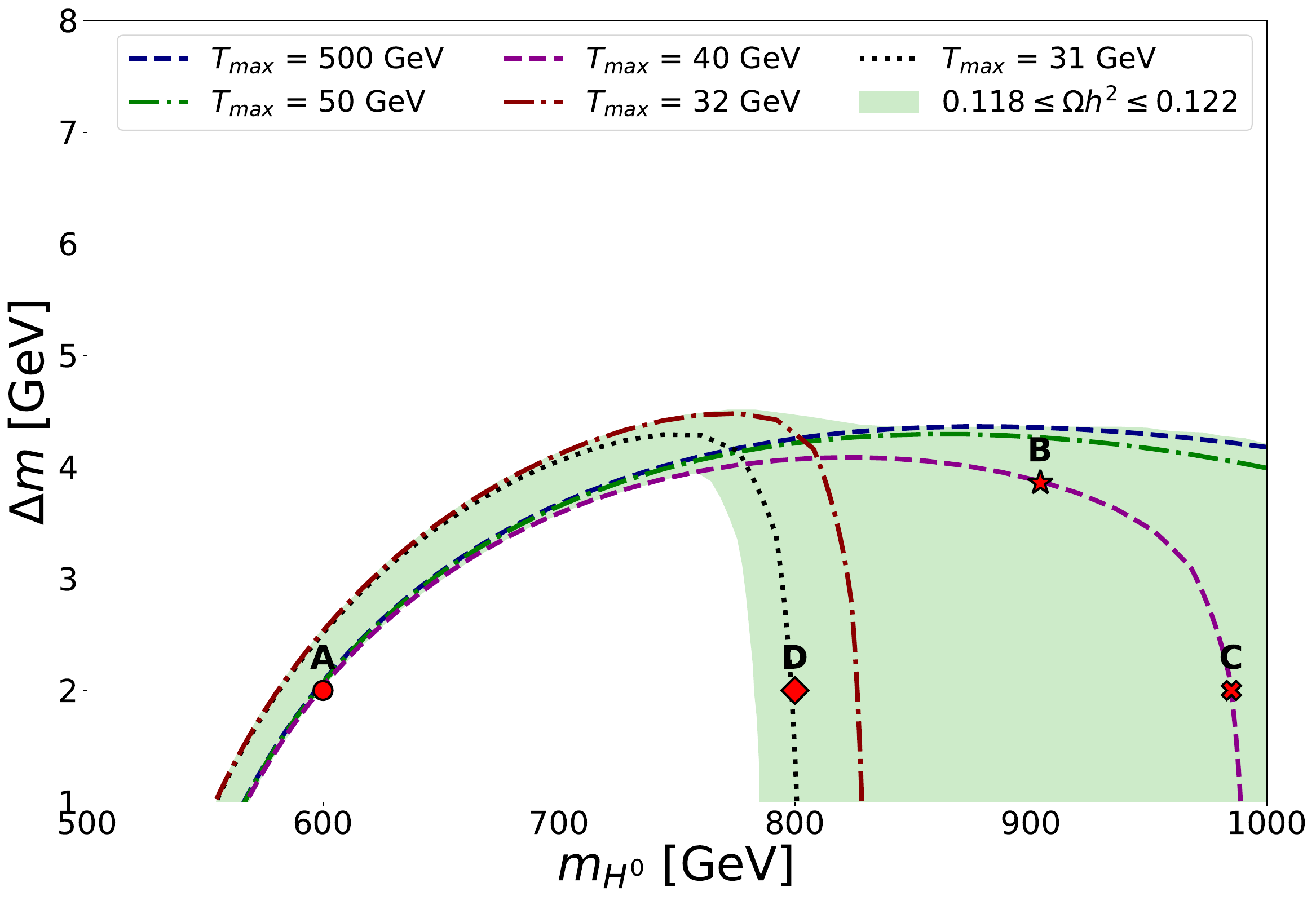}
    \caption{Contours along which the relic density constraint is satisfied assuming a reheating temperature $\Trh = 30$~GeV. The green shaded region represents the union of parameter points satisfying the observed relic-density constraint for at least one value of $\Tmax$ within the considered range. In all cases, the depicted results correspond to the IDM benchmark $\lambda_L = 0$.}
    \label{fig:nonintantaneous}
\end{figure} 
%%%%%%%%%%%%%%%%%%%%%%%%%%%%%%%%%%%%%%%%%%%%%%%%%%%%%%%%%%

With the previous comments in mind, in Fig.~\ref{fig:nonintantaneous} we present the full cosmologically viable parameter space of the IDM in the $(m_{H^0}, \Delta m)$ plane, assuming $\lambda_L = 0$  and $\Trh = 30$~GeV (green shaded area). We, moreover, highlight contours of $\Omega h^2 = 0.12$ for different cosmological evolution scenarios corresponding to $\Tmax = 31$, $32$, $40$, $50$ and $500$~GeV (dotted black, dot-dashed red, dashed purple, dot-dashed green, and dashed blue curves, respectively). The four benchmarks of Fig.~\ref{fig:finitereheatingbenchmarks} are also marked.

The general behavior of the viable parameter space can be understood through the previous discussion. For concreteness, let us first focus on the curve corresponding to $\Tmax = 40$~GeV (dashed purple). For relatively low masses, the DM abundance follows a fairly standard thermal evolution with freeze-out occurring during radiation domination. The shift toward slightly smaller mass splittings relative to the standard cosmological result can be attributed to the modified expansion and entropy evolution during reheating, with residual inflaton decays providing an additional dilution of the DM yield. Once the DM mass becomes sufficiently large, thermalization is no longer possible and $H^0$ freezes-in. These two cases correspond to the top-left and bottom-left benchmarks depicted in Fig.~\ref{fig:finitereheatingbenchmarks}. They are bridged by the situation featured in the top-right panel, in which DM actually freezes out \textit{during} the reheating period. Larger $\Tmax$ values, in turn, imply that equilibration and subsequent freeze-out becomes possible for a larger range of DM masses, which explains the behavior of the different curves for increasing $\Tmax$ and the fact that the freeze-in branch is pushed towards higher values of $m_{H^0}$. In contrast, as $\Tmax$ decreases and approaches $\Trh$, the behavior of the viable parameter space increasingly resembles that of instantaneous reheating. We repeat, however, that we do not expect (neither do we find) the ``instantaneous reheating'' limit to be exactly recovered. In any case, and in comparison with standard cosmology, we find that the viable parameter space is pushed towards scenarios in which DM would have been overabundant. On the other hand, compared to the instantaneous reheating approximation, we find that larger DM masses become accessible, because $\Trh$ no longer acts as a hard cut-off during thermal evolution.

%%%%%%%%%%%%%%%%%%%%%%%%%%%%%%%%%%%%%%%%%%%%%%%%%%%%%%%%%%
\section{Conclusions} \label{sec:conclu}
%%%%%%%%%%%%%%%%%%%%%%%%%%%%%%%%%%%%%%%%%%%%%%%%%%%%%%%%%%
In this work, we have revisited the dark matter (DM) relic abundance in the Inert Doublet Model (IDM), considering cosmological scenarios featuring low-temperature reheating. The IDM provides a particularly interesting framework for this investigation, since its DM candidate has a mass at the electroweak scale or above and interacts with the Standard Model (SM) through electroweak gauge interactions, even when the Higgs-portal coupling vanishes, $\lambda_L = 0$. It therefore possesses the characteristic properties of a Weakly Interacting Massive Particle (WIMP), independently of the cosmological mechanism responsible for generating its abundance.

We have shown that, contrary to the conventional picture, such a particle need not attain chemical equilibrium with the SM plasma. If the temperature of the Universe is sufficiently low, thermal production becomes strongly Boltzmann suppressed, allowing DM to acquire its relic abundance through freeze-in rather than freeze-out, despite its electroweak-strength interactions. In this sense, a genuine WIMP can behave cosmologically as a Feebly Interacting Massive Particle (FIMP).

Within the instantaneous-reheating approximation, we have demonstrated that lowering the reheating temperature $\Trh$ can continuously transform conventional freeze-out into nonequilibrium production, with intermediate regimes in which annihilation backreactions can be relevant. Consequently, regions of the IDM parameter space that would overproduce DM in standard cosmology can reproduce the observed abundance. This result persists when the Higgs-portal coupling is varied within its experimentally allowed range. We have also found that, for a fixed $\Trh$, suppression of thermal production can impose an upper limit on the DM mass that is compatible with the observed abundance.

Going beyond instantaneous reheating, we have shown that the existence of a maximum plasma temperature $\Tmax > \Trh$ allows DM production during the reheating epoch itself. Depending on the hierarchy between the particle masses and the relevant cosmological temperatures, the observed abundance can arise through conventional freeze-out during radiation domination, freeze-out during reheating, or freeze-in. The final relic abundance is then determined by the interplay between thermal production, annihilation processes, the modified expansion history, and entropy injection. In particular, finite-duration reheating can allow heavier DM candidates to reproduce the observed abundance compared to the instantaneous-reheating approximation.

Our results demonstrate that the microscopic properties of a DM candidate do not uniquely determine its cosmological production mechanism. In particular, the presence of electroweak-strength interactions does not guarantee that chemical equilibrium was ever established in the early Universe. The distinction between WIMPs and FIMPs, when understood in terms of their conventional production mechanisms, therefore depends crucially on the underlying cosmological assumptions and parameter regions excluded in conventional freeze-out calculations may become viable under alternative thermal histories. Thus, a WIMP need not be a thermal relic and WIMPs can, from a cosmological standpoint, behave as FIMPs. This observation has important implications for the interpretation of DM searches and for the identification of cosmologically viable regions of the WIMP parameter space. As an example, we could mention the fact that the IDM -- in particular in its very small $\Delta m \sim O(10^2)$~keV regime -- has recently been invoked as a possible explanation of the LZ excess~\cite{LZ:2026axp, Nomura:2026qyq, Wang:2026ytg, Murayama:2026apt}. The results we presented in this paper could be relevant for such interpretations, since they enlarge the model's cosmologically viable parameter space. We postpone such an analysis for future work.

%%%%%%%%%%%%%%%%%%%%%%%%%%%%%%%%%%%%%%%%%%%
\acknowledgments
%%%%%%%%%%%%%%%%%%%%%%%%%%%%%%%%%%%%%%%%%%%
NB received funding from the grants PID2023-151418NB-I00 funded by MCIU/AEI/10.13039 /501100011033/ FEDER and PID2022-139841NB-I00 of MICIU/AEI/10.13039/501100011033 and FEDER, UE.

%%%%%%%%%%%%%%%%%%%%%%%%%%%%%%%%%%%%%%%%%%%%%%%%%%%%%%%%%%
\appendix
\section{The DM yield evolution at low temperature} \label{sec:lowT}
%%%%%%%%%%%%%%%%%%%%%%%%%%%%%%%%%%%%%%%%%%%%%%%%%%%%%%%%%%
In Fig.~\ref{fig:yiedlTrh} we noted that the DM abundance continues evolving non-trivially up to $x \simeq 1000$, beyond which its value saturates. This behavior contradicts naive expectations, which would dictate that the saturation should occur close to $x_{\rm fo}$ in the case of freeze-out and $x_{\rm rh}$ in the case of freeze-in (since, in the low reheating temperature scenarios considered in this paper, freeze-in is effectively UV-dominated).

%%%%%%%%%%%%%%%%%%%%%%%%%%%%%%%%%%%%%%%%%%%%%%%%%%%%%%%%%%
\begin{figure}[t!]
    \def\sepf{0.496}
    \centering
    \includegraphics[width=0.8\columnwidth]{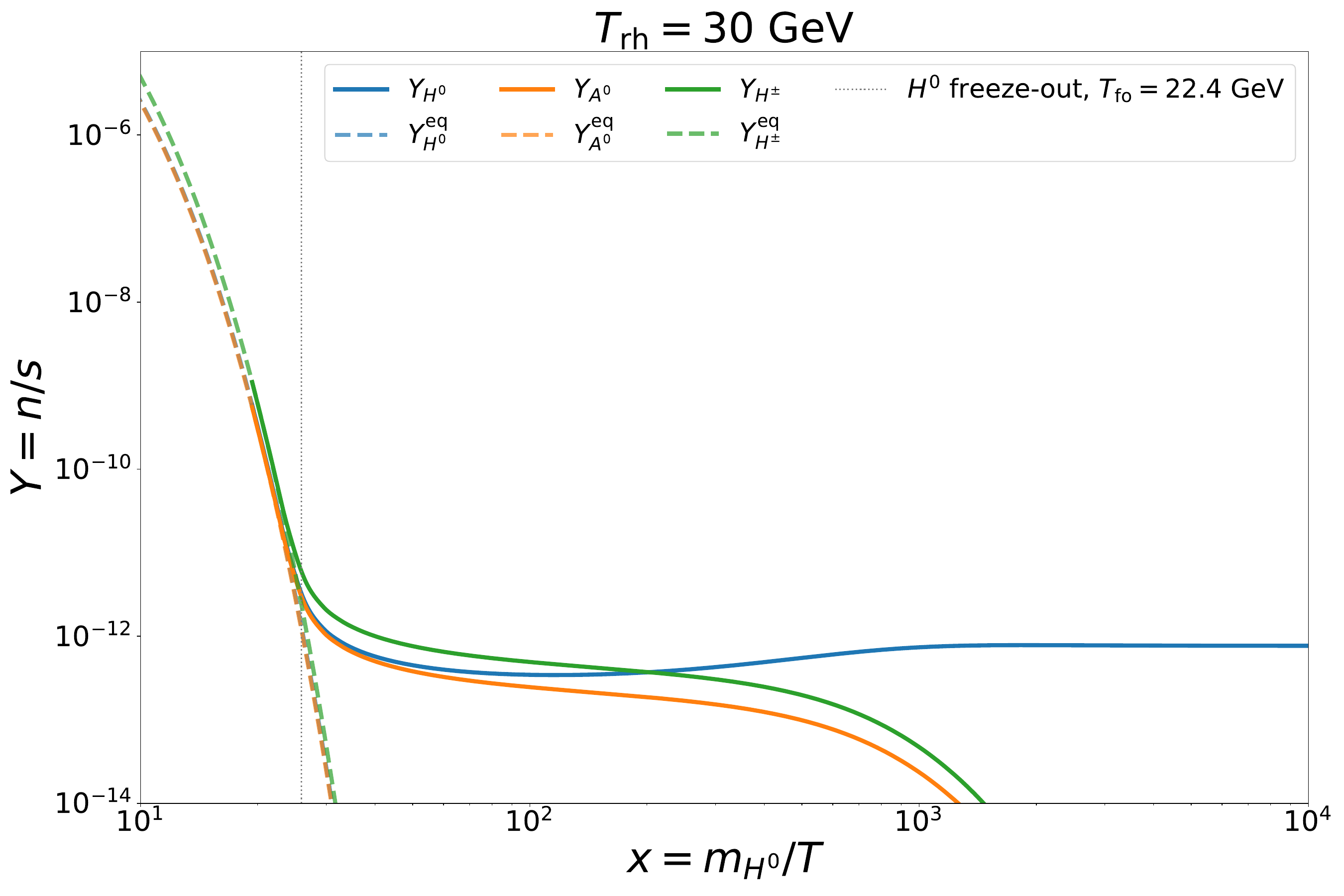}
    \caption{DM yield evolution as a function of $x \equiv m_{H^0}/T$ for a benchmark with $m_{H^0} = 580$~GeV, $\lambda_L = 0$ and $\Delta m = 2$~GeV which satisfies the relic density constraint along with all other relevant bounds, assuming a reheating temperature $\Trh = 30$~GeV.}
    \label{fig:lowTRequilibration}
\end{figure} 
%%%%%%%%%%%%%%%%%%%%%%%%%%%%%%%%%%%%%%%%%%%%%%%%%%%%%%%%%%
%%%%%%%%%%%%%%%%%%%%%%%%%%%%%%%%%%%%%%%%%%%%%%%%%%%%%%%%%%
\begin{figure}[t!]
    \def\sepf{0.496}
    \centering
    \includegraphics[width=0.8\columnwidth]{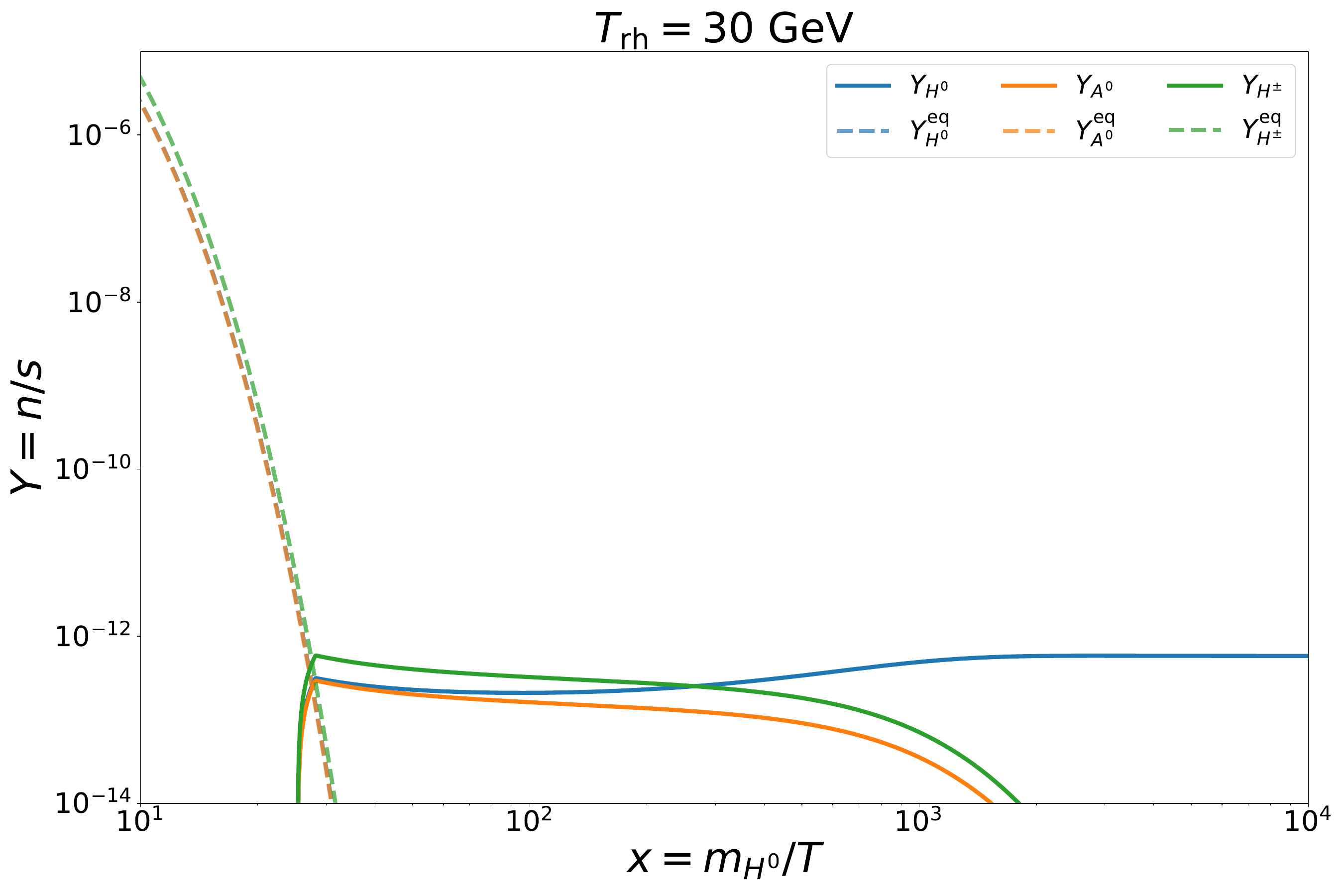}
    \caption{DM yield evolution as a function of $x \equiv m_{H^0}/T$ for a benchmark with $m_{H^0} = 760$~GeV, $\lambda_L = 0$ and $\Delta m = 2$~GeV, assuming a reheating temperature $\Trh = 30$~GeV.}
    \label{fig:lowTRfi}
\end{figure} 
%%%%%%%%%%%%%%%%%%%%%%%%%%%%%%%%%%%%%%%%%%%%%%%%%%%%%%%%%%
In order to understand this effect, in Figs.~\ref{fig:lowTRequilibration} and~\ref{fig:lowTRfi} we show the yield evolution for all three inert sector states $H^0$ (solid blue curves), $A^0$ (solid orange curves) and $H^\pm$ (solid green curves) for two benchmark scenarios $(m_{H^0}, \Delta m) = (580, 2)$ and $(760, 2)$~GeV, respectively. In both cases, $\lambda_L = 0$, the reheating temperature is taken to be $\Trh = 30$~GeV and the dashed curves depict the equilibrium abundances for the three species. The first case corresponds to freeze-out whereas the second one to freeze-in, and the relic density constraint is satisfied in both scenarios.

As we can observe, the late-time increase of the $H^0$ yield is associated with the progressive depletion of the heavier inert states, whose remaining populations are eventually transferred to the stable DM candidate. This behavior may appear surprising, as neither $A^0$ nor $H^\pm$ is particularly long-lived. The origin of this effect lies in conversion processes within the $\mathbb{Z}_2$-odd sector, of the form $H_i + \mathrm{SM} \leftrightarrow H_j + \mathrm{SM}$, where $H_{i,j} \in \{H^0, A^0, H^\pm\}$. Unlike pair annihilations, the rates of these processes are controlled by the abundance of SM particles in the thermal bath and are therefore not directly suppressed by the exponentially small equilibrium densities of the heavy inert states. Consequently, conversions can remain efficient even after chemical equilibrium between the inert sector and the SM plasma has been lost, maintaining relative chemical equilibrium among the different inert particles.

As the temperature decreases below the mass splitting $\Delta m$, conversions into heavier inert states become increasingly Boltzmann suppressed, while the corresponding exothermic processes and decays continue to redistribute the existing inert-sector population toward $H^0$. The heavier states are therefore progressively depleted, leading to the late-time increase of the stable DM yield. Importantly, this evolution does not correspond to the additional production of inert particles from the SM bath but primarily to a redistribution of the previously produced population. Once number-changing reactions become negligible, the total inert-sector yield remains approximately constant, even though the individual abundances can continue evolving.

%%%%%%%%%%%%%%%%%%%%%%%%%
\bibliographystyle{JHEP}
\bibliography{biblio}
%%%%%%%%%%%%%%%%%%%%%%%%%
\end{document}